\documentclass[a4paper,fleqn,usenatbib,useAMS]{mnras}

\usepackage{graphicx}	
\usepackage{amsmath}	
\usepackage{amssymb}	
\usepackage{multicol}        
\usepackage{bm}		
\usepackage{pdflscape}	
\usepackage{subcaption}
\newcommand{\msun}{\ensuremath{\mbox{M}_{\odot}}}

\newcommand{\rsun}{\ensuremath{\mbox{R}_{\odot}}}
\newcommand\0{\phantom{0}}

\usepackage[T1]{fontenc}
\usepackage{ae,aecompl}

\usepackage{newtxtext,newtxmath}
\usepackage{tabularx}

\title[Polarization of $\mu^1$ Sco]{Phase-locked polarization by photospheric reflection in the semidetached eclipsing binary $\mu^1$ Sco}

\author[D. V. Cotton \textit{et al.}]{Daniel V. Cotton$^{1,2,3}$\thanks{Contact e-mail: \href{mailto:daniel.cotton@anu.edu.au}{daniel.cotton@anu.edu.au}}, Jeremy Bailey$^{4}$, Lucyna Kedziora-Chudczer$^{3,4}$, Ain De Horta$^2$\\
$^1$Anglo Australian Telescope, Australian National University, 418 Observatory Road, Coonabarabran, NSW 2357, Australia.\\
{$^2$Western Sydney University, Locked Bag 1797, Penrith-South DC, NSW 1797, Australia.}\\
{$^3$Centre for Astrophysics, University of Southern Queensland, Toowoomba, Queensland. 4350. Australia.}\\
{$^4$School of Physics, UNSW Sydney, New South Wales, 2052, Australia.}
}

\date{Last updated \today; in original form \today}

\pubyear{2020}
\begin{document}
\label{firstpage}
\pagerange{\pageref{firstpage}--\pageref{lastpage}}
\maketitle

\begin{abstract}
We report the detection of phase-locked polarization in the bright ($m_V=2.98-3.24$) semidetached eclipsing binary $\mu^1$ Sco (HD~151890). The phenomenon was observed in multiple photometric bands using two different HIPPI-class (HIgh Precision Polarimetric Instrument) polarimeters with telescopes ranging in size from 35-cm to 3.9-m. The peak-to-trough amplitude of the polarization is wavelength dependent and large, $\sim$700 parts-per-million in green light, and is easily seen with even the smallest telescope. We fit the polarization phase curve with a SYNSPEC/VLIDORT polarized radiative transfer model and a Wilson-Devinney geometric formalism, which we describe in detail. Light from each star reflected by the photosphere of the other, together with a much smaller contribution from tidal distortion and eclipse effects, wholly accounts for the polarization amplitude. In the past polarization in semidetached binaries has been attributed mostly to scattering from extra-stellar gas. Our new interpretation facilitates determining masses of such stars in non-eclipsing systems.
\end{abstract}

\begin{keywords}
binary stars; stars: individual: $\mu^1$ Sco; methods: polarimetry
\end{keywords}



\section{Introduction}
\label{sec:intro}

\subsection{Phase-locked polarization in early-type binaries}

Polarization phase-locked to the orbital period in an early-type binary star was first seen in the eclipsing system $\beta$~Lyr in the 1960s \citep{Serkowski65, Appenzeller67}\footnote{$\beta$~Lyr was the first claimed polarimetric variable \citep{Ohman34}, but it wasn't until \citet{Shakhovskoi62}'s measurements were reproduced by other observers (refs within \citealp{Serkowski65}) that its variability was accepted.}. Two decades earlier \citet{Chandrasekhar46} had shown that a net polarization would be produced in a binary system as an eclipsing star breaks the disc symmetry of its companion\footnote{Referred to as either the \textit{Chandrasekhar Effect} or the \textit{Eclipse Effect}.}, but that an isolated spherical star is unpolarized. Yet, in $\beta$~Lyr the polarization varied smoothly outside of eclipse. \citet{Shakhovskoi64} proposed that electron scattering from the extended co-rotating gaseous envelopes in the system was responsible, this mechanism quickly became favoured, with other versions differing only in the configuration of the gaseous material \citep{Rucinski66, Appenzeller67}. When other polarimetric binaries were discovered (e.g. \citealp{Pfeiffer73, Rudy76, Kemp77}), models based on the same basic mechanism -- gaseous envelopes or streams of co-rotating electron (Thomson) scattering gas between the stars -- were developed to fit the data \citep{Brown78, Rudy78}.

Because polarimetry includes orientation information, it can resolve orbital elements that spectroscopy and photometry together cannot. This was first realised by \citet{Shakhovskoi66} who, after subtracting interstellar polarization, determined the position angle of the line-of-nodes in a number of eclipsing systems \citep{Shakhovskoi69}. In the process of developing models to describe polarization in binary systems by extra-stellar gas, \citet{Rudy78} and \citet{Brown78} independently demonstrated that polarimetry could be used to determine the system's orbital inclination, and therefore the component masses in a binary system. It is only possible to determine the component masses of a spectroscopic binary when the inclination is known by other means. Usually astronomers rely on photometry from eclipsing systems for this information. 

In the model of \citet{Rudy78}, phase-locked polarization, with a pattern repeated twice per orbit, arises principally by single scattering from optically thin extra-stellar material. So long as there is no eclipse of the scattering material, its distribution is symmetric about the orbital plane, and the photometric variability not significant, the inclination of the system is derived from the eccentricity of the elliptical trace of the polarization in a \mbox{Q-U} diagram. Though founded on the same assumptions, the \textit{\mbox{BMcLE} model} \citep{Brown78} is more sophisticated; it is a complete Fourier analysis that allows the orbital inclination and moment integrals of the density distribution to be determined, even for envelopes with no symmetry about the orbital plane. In principle this allows calculation of the scattering mass \citep{Koch89} and optical depth \citep{Aspin82} and permits different scattering geometries to be distinguished, though in practice there is often degeneracy (e.g. \citealp{Berdyugin18}). 

With few exceptions, the polarimetric binaries characterised over the subsequent 40~years either consisted of interacting stars or those with extended envelopes. In summarising the sources of polarization in binary systems \citet{Pfeiffer77} state ``\textit{With one exception (U~Oph\footnote{Despite a large number of studies, the polarigenic mechanism(s) operating in the U~Oph system are not certain, though most recently mass-loss via a strong stellar wind, and eruptive mass outflows due to asynchronous rotation of the components have been favoured \citep{Eritsian98}.}), un-evolved binaries do not exhibit intrinsic polarization}''. Overwhelmingly the preferred mechanism to explain the polarization in these systems was electron scattering from gas shells or streams of gas entrained between the companions (e.g. \citealp{Pfeiffer77, McLean80}; refs. within \citealp{Clarke10}). Typically data variability beyond the formal errors is naturally explained by irregularities in the flow of the transfer material \citep{Koch89}. More recently \citet{Berdyugin16, Berdyugin18} explored different configurations and densities of the entrained gas in the semi-detached binaries HD~48099 and $\lambda$~Tau. Instead of the \textit{\mbox{BMcLE} model}, they used a numerical scattering code, which allows the inclusion of specific scattering geometries with eclipse effects to provide tighter constraints on the distribution of the scattering material. Such models, given sufficient instrumental precision, are good at determining binary inclination, which is fairly insensitive to the gas geometry \citep{Berdyugin18}. The position angle of the line of nodes, however, is sensitive to geometry \citep{Berdyugin16}. This conventional interpretation of binary polarization, implies that the ability of polarimetry to determine inclinations is limited to interacting stars.

In addition to the \textit{Chandrasekhar effect} and scattering from extra-stellar material, there are other potential sources of phase-locked polarization in binaries. In considering the polarigenic behaviour of the semidetached u~Her\footnote{68~Her.} system, \citet{Rudy77} also considered (i) photospheric reflection, (ii) tidal distortion, and (iii) asymmetric temperature distribution, i.e. as induced by the \textit{\mbox{reflection effect}} -- a misnomer referring to the effects of the heating of one star by the other \citep{Eddington26}. An inceptive theoretical radiative transfer study of polarization arising from the reflection effect had been made by \citet{Collins74}, and this and tidal distortion were ruled out on the basis of the low temperature and small distortion of the companion. Instead, while not ruling out extra-stellar material, \citet{Rudy77} favoured photospheric reflection by the primary from the secondary for u~Her. This mechanism was also considered likely for at least part of the out-of-eclipse polarization of Algol by \citet{Kemp81}, who made order of magnitude estimates for it. Tidal distortion effects have since been modelled for the visible counterpart of the X-ray source Cyg~X-1 \citep{Bochkarev86, Dolan92}, and lately for the general case by \citet{Harrington}, who shows the effects to be relatively small. \citet{Bott18} have examined the effect on a hot-Jupiter exoplanet with the same finding.

After decades of searching, observations showing the Chandrasekhar effect were reported for Algol by \citet{Kemp83}, as a secondary effect to scattering by entrained gas. Algol remains the only star the Chandrasekhar effect has been observed in. Photospheric reflection was claimed as a component of the observed polarization in HR~5110 \citep{Barbour81} and V373~Cas \citep{Berdyugin98}, but no other convincing examples of other dominant mechanisms were seen until \citet{Berdyugin99} studied the ellipsoidal binary LZ~Cep. Spectroscopic data limited the amount of extra-stellar gas that could be in the system, and this was insufficient to account for the polarization through a scattering mechanism. The polarization amplitude was also greater than could be accounted for by tidal distortion or heating. They concluded that photospheric reflection was dominant. Their simple numerical model showed reflected light to be about 3~per~cent of the integrated flux, but was unable to explain the wavelength dependence of the polarization. Subsequently photospheric reflection was also considered a plausible mechanism for polarization in the HD~48099 system \citep{Berdyugin16}.

Recently we \citep{Bailey19} studied the detached ellipsoidal binary Spica. It displays the same double-peaked polarization curve typical of all the aforementioned binaries. It does not have any entrained gas, and a modified version of our polarized radiative transfer code \citep{Cotton17} revealed that the majority of the polarization is accounted for by photospheric reflection, with a smaller amount ascribed to tidal distortion. Importantly, the Spica model constructed from its known stellar and orbital parameters, required no fitting to match the polarization amplitude -- only the interstellar offsets and the line-of-nodes was fit. Interestingly the modelled dependence on wavelength \citep[][Supplementary Information]{Bailey19} is reminiscent of that observed for LZ~Cep \citep{Berdyugin99}. 

The polarization amplitude in the Spica system is small compared to other known polarimetric binaries, about 200 parts-per-million (ppm). However, our models \citep{Cotton17, Bailey19} indicate that the amplitude will be larger in atmospheres that are hotter and/or have lower gravity -- which is a noted trend in such systems \citep{Koch89}. Photospheric reflection may therefore account for the polarization in more binary systems than presently thought.

\subsection{\texorpdfstring{$\mu^1$ Sco}{mu1 Sco}}
\label{sec:mu1Sco_intro}

$\mu^1$~Sco (Xamidimura, HR~6247, HD~151890, HIP~82514) is one of the brightest ($m_V=2.98$, $B-V=0.16$) Algol-type eclipsing binaries; it has a primary eclipse depth, $\Delta m$, of 0.28 \citep{vanAntwerpen10}. It was only the third star to be recognised as a spectrocopic binary \citep{Pickering96}. Drawing on decades of radial velocity and photometry, and their own observations, \citet{vanAntwerpen10} determine the orbit to be circular with an inclination of 65.4~$\pm$~1.0$^\circ$ and a semi-major axis of 12.90~$\pm$~0.04~$R_\odot$, and a period of 1.4462700(5)~days (identical to \citealp{Maury20} who was the first to study it extensively), with a primary photometric minimum at ephemeris HJD~2449534.17700(9). The mass of the B1.5~V primary is found to be 8.49~$\pm$~0.05~$M_\odot$ and the B8--B3 secondary 5.33~$\pm$~0.05~$M_\odot$, the radius of the primary is 4.07~$\pm$~0.05~$R_\odot$ and the secondary 4.38~$\pm$~0.05~$R_\odot$ \citep{vanAntwerpen10} -- which makes it considerably oversized for its mass \citep{Budding15}. A full list of parameters pertinent to the current work is given later in section \ref{sec:modelling}.

$\mu^1$~Sco is a semi-detached binary where the secondary is close to or just filling its Roche lobe \citep{vanAntwerpen10, Budding15}. According to \citet{Budding15} the secondary's over-sized character implies on-going late-state interactive stellar evolution, believed to be of the `Case~A' type (i.e. it's still on the main sequence), though the primary is little effected. They further note that the system is atypical for an Algol-type binary in that the components are especially close, their masses high, and the primary particularly hot. Though gas flows in Algol-type binaries are expected to be faint, \citet{vanAntwerpen10} suggest that it may be possible to detect the gas streams between the stars in $\mu^1$~Sco spectroscopically using Doppler tomography.

The most recent new data on $\mu^1$~Sco was collected by \citet{Handler13}, who made multi-band photometric observations, including the first such in \textit{u} band. This ultraviolet data required inclusion of the reflection effect to fit the observations within the observational errors. \citet{Handler13} were interested in the system because the primary lies in the $\beta$~Cep instability strip and the secondary in the domain of slowly pulsating B-type (SPB) stars. Which is pertinent to any polarimetric study in view of \citet{Odell79}'s prediction of polarization from non-radial pulsations. After removing the binary light curve solution from their data \citet{Handler13} are left with two non-white frequencies at 0.123 and 8.07~c/d with an amplitude of around 3~mmag, but a S/N ratio less than is required for detection. 

Rather surprisingly, $\mu^1$~Sco has only been observed polarimetrically in a single study. As part of a survey of early-type stars, \citet{Serkowski70} observed the system just twice in the \textit{B} band at JD 2439949.19 and 2439954.24, measuring $p=$ 3634~ppm, $PA=$ 21$^\circ$; and $p=$ 4094~ppm, $PA=$ 18$^\circ$ respectively\footnote{\label{fn:pol_mag}The data in the source material is given in polarization magnitudes and has been converted to fractional polarization.} (the formal error is given as 100~ppm, but is almost certainly somewhat larger). \citet{Serkowski70} noted the binary phases of the measurements, but otherwise made no special mention of the system, and did not include it amongst the early-type stars he considered intrinsically polarized in his conclusions.

$\mu^1$~Sco has a wide (347$\arcsec$) companion, $\mu^2$ Sco (Pipirima, HR~6252, HD~151985, HIP~82545), which has a spectral type of B2\,IV and a magnitude of $m_V=3.6$. Despite being labelled as ``Variable'' in SIMBAD\footnote{The origin of this determination is not clear.} $\mu^2$~Sco is not considered so by most observers, and is frequently used as a constant check star in observations of $\mu^1$~Sco \citep{vanAntwerpen10, Handler13}. It was the subject of two unfiltered\footnote{The detector used was a 1P21 photomultiplier tube, which depending on the model and operating temperature has a peak efficiency at $\sim$450~nm, but has some sensitivity between 300 and 650~nm.} polarimetric observations by \citet{Smith56} giving an average $p=$ 4605~$\pm$~$\sim<$1302~ppm$^{\ref{fn:pol_mag}}$, $PA=$ 25~$\pm$~4$^\circ$. That these values are the same as those of $\mu^1$~Sco within error suggests that the majority of the polarization is interstellar in origin.

\bigbreak

In  this  paper  we  report  polarimetric observations  of $\mu^1$~Sco (and $\mu^2$~Sco) in section \ref{sec:observations}. In section \ref{sec:modelling} we describe polarization models, which have been developed using a similar approach to that adopted in \citet{Bailey19}. In section \ref{sec:results} the model results are shown and compared to the observations. We discuss the implications of the results in section \ref{sec:discussion} and conclude in section \ref{sec:conclusions}.

\section{Observations}
\label{sec:observations}

\begin{table*}
\caption{Summary of Mini-HIPPI and HIPPI-2 Observing Runs}
\label{tab:runs}
\tabcolsep 3 pt
\begin{tabular}{cl|ccrccc|rrrrr|rr|r}
\hline
\hline
Run & JD Range      & \multicolumn{6}{c|}{Telescope and Instrument Set-Up$^a$}                                      &   \multicolumn{5}{c|}{Observations$^b$}                                                                   &   \multicolumn{2}{c|}{TP$^c$}    & \multicolumn{1}{c}{$\sigma_{PA}^d$}\\
    &  2450000+     &   Instr.      &   Tel.   &\multicolumn{1}{c}{f/}   &   Fil.   &   Mod.    &   Ap. ($\arcsec$) &   n   &   Exp. (s)            &  Dwell (s)            &   $\lambda_{eff}$ (nm)    &  Eff ($\%$)           &    \multicolumn{1}{c}{$q$ (ppm)} &   \multicolumn{1}{c|}{$u$ (ppm)} & \multicolumn{1}{c}{($^{\circ}$)}\\
\hline
MC1& 7871--7898       &  M-HIPPI      &   UNSW    &11\phantom{.0*}& Cl                &   MT      &  58.9             &  34   &   800\phantom{.$^0$}   & 1595$\pm$167          &   453.1$\pm$2.0           & 74.0$\pm$0.8          &  $-$69.8$\pm$2.9 &  $-$10.9$\pm$2.9  & 0.23\\ 
MC2& 8017--8049       &  M-HIPPI      &   UNSW    &11\phantom{.0*}& Cl                &   MT      &  64.3             &  31   &   800\phantom{.$^0$}   & 1746$\pm$301          &   456.1$\pm$2.9           & 75.2$\pm$1.1          &  $-$73.6$\pm$8.8 & $-$20.7$\pm$8.6  & 0.52\\ 
AC1& 8153       &  HIPPI-2      &   AAT &   15\phantom{.0*} & Cl                &   BNS-E3      &  16.8             &   2   &   400$^E$              &  618$\pm$233          &   454.2$\pm$0.3           & 67.1$\pm$0.1          & $-$185.6$\pm$1.0 &     8.6$\pm$0.9  & \multicolumn{1}{c}{-}\\
AC2& 8154       &  HIPPI-2      &   AAT &   15\phantom{.0*} & Cl                &   BNS-E3      &  16.8             &   1   &   200\phantom{.$^0$}   &  448\phantom{$\pm$000}&   454.0\phantom{$\pm$0.0} & 67.0\phantom{$\pm$0.0}& $-$175.2$\pm$0.8 &    14.0$\pm$0.8  & \multicolumn{1}{c}{-}\\
AC3& 8211--8216        &  HIPPI-2      &   AAT &   8*\phantom{.0}  & Cl                &   BNS-E3      &  11.9             &   6   &   160\phantom{.$^0$}   &  405$\pm$\phantom{0}43          &   457.6$\pm$0.1           & 69.7$\pm$0.0          &    113.3$\pm$0.7 &     3.9$\pm$0.9  & 0.26\\
MC3& 8343--8344       &  M-HIPPI      &   UNSW    &11\phantom{.0*}& Cl                &   MT      &  64.3             &   6   &   400\phantom{.$^0$}   & 1434$\pm$615          &   452.5$\pm$1.4           & 73.7$\pm$0.6          &  $-$56.4$\pm$1.9 &    16.2$\pm$1.9  & 0.54\\ 
AC4& 8360--8361        &  HIPPI-2      &   AAT &   8*\phantom{.0}  & Cl                &   BNS-E7      &  11.9             &   9   &   320$^E$              &  650$\pm$159          &   459.0$\pm$0.7           & 50.4$\pm$0.3          &  $-$10.1$\pm$0.9 &     3.8$\pm$0.9  & 0.86\\
AC5& 8590--8605        &  HIPPI-2      &   AAT &   15\phantom{.0*} & Cl                &   ML-E1      &  12.7             &   7   &   160$^E$              &  602$\pm$353          &   449.6$\pm$1.6           & 80.7$\pm$0.4          &  $-$14.2$\pm$0.8 &  $-$3.7$\pm$0.7  & 0.27\\
\hline
AG1& 8593--8605        &  HIPPI-2      &   AAT &   15\phantom{.0*} & $g^{\prime}$      &   ML-E1      &  12.7             &   8   &   160$^E$             &  409$\pm$\phantom{0}40 &   458.1$\pm$0.6           & 92.9$\pm$0.0          & $-$14.2$\pm$0.8 &   $-$2.6$\pm$0.7  & 0.27\\
WG1& 8662--8665        &  HIPPI-2      &   WSU &   10.5*           & $g^{\prime}$      &   ML-E1      &  58.9             &   4   &   800\phantom{.$^0$}   & 1256$\pm$160          &   458.6$\pm$1.0           & 93.0$\pm$0.0          &  $-$4.6$\pm$3.8 &      3.1$\pm$3.9  & 0.09\\
AG2& 8669--8678     &  HIPPI-2      &   AAT &   15\phantom{.0*} & $g^{\prime}$      &   ML-E1      &  12.7             &  11   &   320$^E$             &  650$\pm$118           &   458.1$\pm$0.5           & 92.9$\pm$0.1          & $-$14.5$\pm$1.1 &      5.0$\pm$1.0  & 0.08\\
WG2& 8704--8707       &  HIPPI-2      &   WSU &   10.5*           & $g^{\prime}$      &   ML-E1      &  58.9             &   6   &   800\phantom{.$^0$}   & 1293$\pm$183          &   458.6$\pm$0.5           & 93.0$\pm$0.0          & $-$32.6$\pm$3.9 &     11.6$\pm$3.8  & 0.05\\
WG3& 8718--8720       &  HIPPI-2      &   WSU &   10.5*           & $g^{\prime}$      &   ML-E1      &  58.9             &   5   &   800\phantom{.$^0$}   & 1145$\pm$\phantom{0}17&   459.2$\pm$0.9           & 93.0$\pm$0.0          & $-$32.6$\pm$3.9 &     11.6$\pm$3.8  & 0.18\\
WG4& 8759        &  HIPPI-2      &   WSU &   10.5*           & $g^{\prime}$      &   ML-E1      &  58.9             &   1   &   800\phantom{.$^0$}   & 1135\phantom{$\pm$000}&   459.6\phantom{$\pm$0.0} & 93.0\phantom{$\pm$0.0}& 1.8$\pm$2.2 &     1.2$\pm$2.1  & 0.44\\
WG5& 8779--8782 &  HIPPI-2      &   WSU &   10.5*           & $g^{\prime}$      &   ML-E1      &  58.9             &   5   &   800\phantom{.$^0$}   & 1131$\pm$\phantom{0}11&   461.7$\pm$1.1 & 93.0$\pm$0.0    & 1.8$\pm$2.2 &     1.2$\pm$2.1  & 0.61\\
\hline
AR1& 8590--8602        &  HIPPI-2      &   AAT &   15\phantom{.0*} & $r^{\prime}$      &   ML-E1      &  12.7             &  10   &   320$^E$             &  690$\pm$118 &   601.2$\pm$0.0           & 62.0$\pm$0.0          & $-$0.7$\pm$2.3 &      14.1$\pm$2.4  & 0.40\\
AR2& 8669--8678        &  HIPPI-2      &   AAT &   15\phantom{.0*} & $r^{\prime}$      &   ML-E1      &  12.7             &   7   &   480$^E$             &  904$\pm$368           &   601.2$\pm$0.1           & 62.0$\pm$0.0          & $-$7.0$\pm$6.9 &   $-$23.9$\pm$7.1  & \multicolumn{1}{c}{-}\\
\hline
\hline
\end{tabular}
\begin{flushleft}
Notes: \textbf{*} Indicates use of a 2$\times$ negative achromatic lens effectively making the focii f/16 and f/21 at the AAT and WSU telescope respectively. \textbf{$^E$} Indicates there are a small number of exceptions to the listed dwell time. \textbf{$^a$} A full description along with transmission curves for all the components and modulation characterisation of each modulator (Mod) in the specified performance era (En) can be found in \citet{Bailey20}. \textbf{$^b$} The dwell time (Dwell), effective wavelength ($\lambda_{eff}$) and modulation efficiency (Eff.) for the set of observations are described as the median $\pm$ the standard deviation. \textbf{$^c$} TP has been calibrated as the mean of observations of the following low polarization standards made during each run, MC1: 4$\times$ HD~2151, 14$\times$ HD~48915, 6$\times$ HD~102647, 1$\times$ HD~102870, 1$\times$ HD~140573; MC2: 2$\times$ HD~2151, 1$\times$ HD~48915, 2$\times$ HD~128620J; MC3: 9$\times$ HD~48915, 3$\times$ HD~128620J;  AC1, AC2, AC3 \& AC4: given in \citet{Bailey20} as runs 2018FEB-B, 2018FEB-D, 2018MAR \& 2018AUG respectively; AC5: 3$\times$ HD~48915, 3$\times$ HD~102647, 1$\times$ HD~140573; AG1: 4$\times$ HD~48915, 3$\times$ HD~102647, 2$\times$ HD~128620J, 1$\times$ HD~140573; AG2: 3$\times$ HD~2151, 3$\times$ HD~102647; WG1:  4$\times$ HD~102647, 4$\times$ HD~128620J;  WG2 \& WG3: 1$\times$ HD~2151, 1$\times$ HD~10700, 5$\times$ HD~128620J, 1$\times$ HD~140573, WG4 \& WG5: 4$\times$ HD~48915, 2$\times$ HD~128620J. \textbf{$^d$} Shown is the standard deviation of PA calibration observations with respect to their literature values (see \citealp{Bailey20}). PA has been calibrated as the mean of observations of the following high polarization standards made in SDSS $g^{\prime}$ or Clear during each run, MC1: 2$\times$ HD~84810, 1$\times$ HD~147084; MC2: 1$\times$ HD~147084, 1$\times$ HD~187929; MC3: 3$\times$ HD~147084, 2$\times$ HD~149757; AC1: 1$\times$ HD~80558; AC2: 1$\times$ HD~80558; AC3: 1$\times$ HD~80558, 1$\times$ HD~111613, 1$\times$ HD~147084, 1$\times$ HD~187929; AC4: 3$\times$ HD~147084, 3$\times$ HD~160529, 5$\times$ HD~187929 (which were then modified according to $\Delta PA = -4.96338\times10^{-2}\lambda_{eff}^2 + 4.278\lambda_{eff} - 915.5$ which is a fit to the standard observations made in SDSS $g^{\prime}$, 500SP -- 1$\times$ HD~147084, 2$\times$ HD~160529, 2$\times$ HD~187929 -- and 425SP -- 1$\times$ HD~147084, 2$\times$ HD~160529, 2$\times$ HD~187929); AC5, AG1: 1$\times$ HD~80558, 1$\times$ HD~111613, 2$\times$ HD~147084, 1$\times$ HD~187929; WG1: 10$\times$ HD~147084, 1$\times$ HD~154445; AG2: 4$\times$ HD~147084, 1$\times$ HD~154445; WG2: 1$\times$ HD~147084, 1$\times$ HD~187929;  WG3: 1$\times$ HD~147084, 1$\times$ HD~187929; WG4: 1$\times$ HD~147084, 1$\times$ HD~187929; WG5: 1$\times$ HD~160529, 2$\times$ HD~187929. The initial PA determinations for AR1 \& AR2 are those for AG1 \& AG2, a correction was carried out for SDSS $r^{\prime}$ based on a smaller number of standard observations made in the same band, AR1: 1$\times$ HD~147084, 1$\times$ HD~187929, (offset 1.58$^{\circ}$); AR2: 1$\times$ HD~147084 (offset $-$1.39$^{\circ}$), tabulated values are for these observations.\\
\end{flushleft}
\end{table*}

\begin{figure*}
\includegraphics[width=\textwidth,trim={0.3cm 0.0cm 0.8cm 0cm},clip]{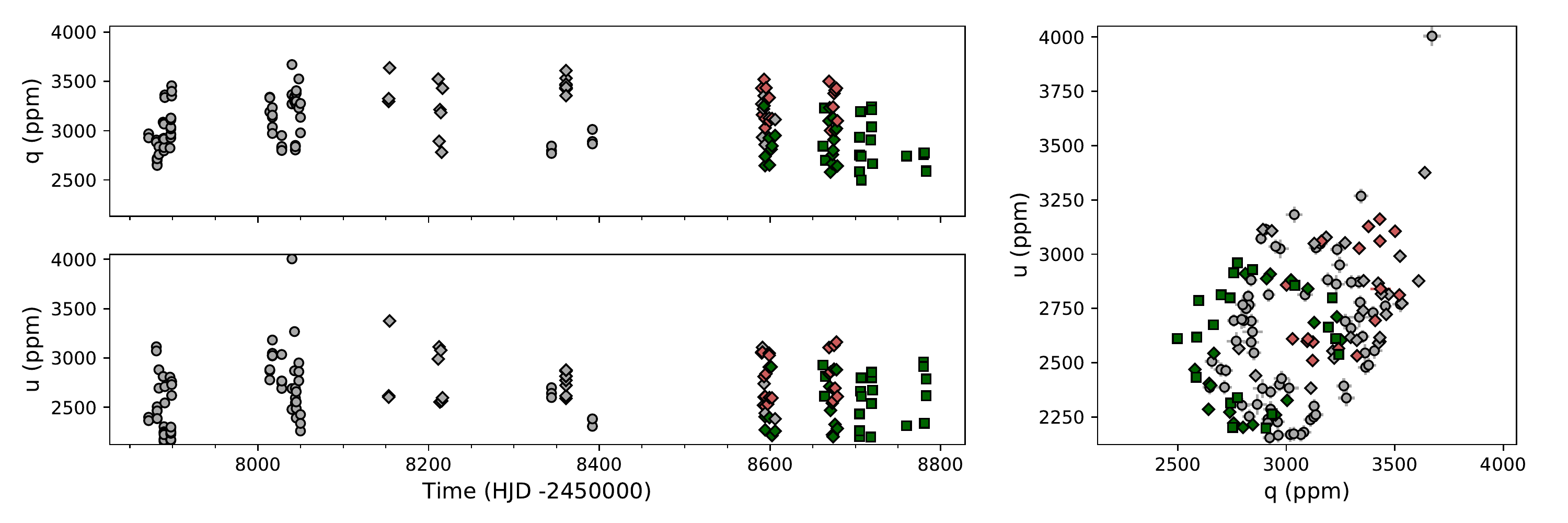}
\caption{All polarimetric observations of $\mu^1$ Sco (after subtraction of TP) in time series for $q=Q/I$ Stokes (top left), $u=U/I$ Stokes (bottom left), and as a Q-U diagram (right). Clear observations are in grey, SDSS $g^{\prime}$ in dark green, SDSS $r^{\prime}$ in light red, UNSW data are circles, WSU squares and AAT diamonds.}
\label{fig:observations}
\end{figure*}

\begin{table*}
\caption{High precision polarimetric observations of $\mu^1$ Sco.}
\label{tab:observations}
\fontsize{7.4pt}{8.25pt}\selectfont
\tabcolsep 4.4pt
\centering
\begin{tabular}{|ccccccc||ccccccc|}
 \hline
 HJD & Phase & Run    &   $\lambda_{eff}$ & Eff & $q$ & $u$ &  HJD & Phase   & Run    &   $\lambda_{eff}$ & Eff & $q$ & $u$ \\
 2450000+ &&& (nm) & (\%) & (ppm) & (ppm) & 2450000+ &&& (nm) & (\%) & (ppm) & (ppm) \\
\hline
7871.99062 & 0.04637 & MC1 & 456.2 & 75.2 & 2968.5 $\pm$  31.4 & 2399.6 $\pm$  31.9 & 8344.00325 & 0.41187 & MC3 & 451.7 & 73.4 & 2769.2 $\pm$  43.6 & 2599.6 $\pm$  42.5 \\
7872.01282 & 0.06172 & MC1 & 454.4 & 74.5 & 2927.6 $\pm$  31.1 & 2365.4 $\pm$  30.8 & 8360.94916 & 0.12885 & AC4 & 458.2 & 50.0 & 3425.4 $\pm$  10.6 & 2590.2 $\pm$ \08.6 \\
7881.03107 & 0.29725 & MC1 & 452.4 & 73.6 & 2905.7 $\pm$  29.2 & 3114.1 $\pm$  29.1 & 8360.95832 & 0.13518 & AC4 & 458.4 & 50.1 & 3431.3 $\pm$  10.5 & 2616.1 $\pm$  13.8 \\
7881.05098 & 0.31101 & MC1 & 451.6 & 73.3 & 2883.6 $\pm$  29.3 & 3071.8 $\pm$  28.3 & 8360.96612 & 0.14058 & AC4 & 458.6 & 50.2 & 3460.6 $\pm$  21.7 & 2722.6 $\pm$  17.6 \\
7881.93520 & 0.92239 & MC1 & 459.6 & 76.6 & 2698.6 $\pm$  32.5 & 2469.0 $\pm$  32.8 & 8360.97692 & 0.14804 & AC4 & 458.9 & 50.3 & 3532.6 $\pm$  31.3 & 2773.2 $\pm$  21.7 \\
7881.95598 & 0.93676 & MC1 & 457.1 & 75.6 & 2657.3 $\pm$  33.7 & 2506.2 $\pm$  33.7 & 8360.98402 & 0.15295 & AC4 & 459.0 & 50.4 & 3423.5 $\pm$  17.5 & 2866.3 $\pm$  20.3 \\
7881.97747 & 0.95162 & MC1 & 455.0 & 74.8 & 2647.0 $\pm$  32.0 & 2385.9 $\pm$  32.3 & 8360.99097 & 0.15776 & AC4 & 459.3 & 50.5 & 3472.0 $\pm$  25.2 & 2814.6 $\pm$  23.7 \\
7881.99841 & 0.96610 & MC1 & 453.7 & 74.2 & 2721.2 $\pm$  32.7 & 2463.9 $\pm$  30.9 & 8360.99798 & 0.16261 & AC4 & 459.6 & 50.7 & 3438.8 $\pm$  11.8 & 2816.7 $\pm$  16.7 \\
7882.02274 & 0.98292 & MC1 & 452.6 & 73.7 & 2715.4 $\pm$  33.5 & 2386.7 $\pm$  31.6 & 8361.00492 & 0.16740 & AC4 & 459.8 & 50.8 & 3356.0 $\pm$ \08.3 & 2877.5 $\pm$  10.2 \\
7884.00376 & 0.35267 & MC1 & 453.2 & 74.0 & 2837.6 $\pm$  30.5 & 2881.1 $\pm$  30.5 & 8361.01208 & 0.17235 & AC4 & 460.1 & 51.0 & 3610.3 $\pm$  11.5 & 2876.7 $\pm$ \09.7 \\
7884.05100 & 0.38533 & MC1 & 451.4 & 73.2 & 2758.5 $\pm$  30.4 & 2694.0 $\pm$  30.0 & 8391.89104 & 0.52311 & MC3 & 452.5 & 73.7 & 2890.4 $\pm$  43.4 & 2381.2 $\pm$  43.2 \\
7888.98350 & 0.79583 & MC1 & 453.5 & 74.1 & 3086.7 $\pm$  35.1 & 2812.7 $\pm$  32.7 & 8391.91995 & 0.54310 & MC3 & 453.9 & 74.3 & 2865.9 $\pm$  45.5 & 2307.9 $\pm$  47.2 \\
7889.05586 & 0.84586 & MC1 & 451.0 & 73.1 & 2918.3 $\pm$  30.1 & 2812.7 $\pm$  31.2 & 8391.93471 & 0.55331 & MC3 & 454.8 & 74.7 & 3012.8 $\pm$  44.1 & 2383.2 $\pm$  44.1 \\
7889.94768 & 0.46249 & MC1 & 455.8 & 75.1 & 2795.5 $\pm$  34.7 & 2303.9 $\pm$  32.5 & 8590.30862 & 0.71574 & AC5 & 449.1 & 80.6 & 3271.1 $\pm$ \04.4 & 3052.0 $\pm$ \04.5 \\
7889.96742 & 0.47614 & MC1 & 454.3 & 74.5 & 2829.2 $\pm$  31.9 & 2252.3 $\pm$  32.7 & 8590.31604 & 0.72087 & AR1 & 601.2 & 62.0 & 3431.8 $\pm$  16.6 & 3060.3 $\pm$  16.6 \\
7889.98848 & 0.49070 & MC1 & 453.2 & 74.0 & 2915.4 $\pm$  31.8 & 2239.8 $\pm$  32.2 & 8591.17797 & 0.31683 & AC5 & 448.6 & 80.4 & 2933.0 $\pm$ \04.5 & 3107.2 $\pm$ \04.3 \\
7890.00577 & 0.50266 & MC1 & 452.4 & 73.6 & 2923.5 $\pm$  31.7 & 2154.4 $\pm$  31.4 & 8591.18550 & 0.32204 & AR1 & 601.1 & 62.0 & 3162.1 $\pm$  15.5 & 3059.8 $\pm$  15.1 \\
7890.02198 & 0.51387 & MC1 & 451.8 & 73.4 & 3081.1 $\pm$  33.4 & 2179.8 $\pm$  31.7 & 8592.31415 & 0.10243 & AC5 & 449.3 & 80.6 & 3220.8 $\pm$ \05.6 & 2521.3 $\pm$ \05.1 \\
7890.03894 & 0.52559 & MC1 & 451.3 & 73.2 & 3067.3 $\pm$  32.2 & 2168.8 $\pm$  31.6 & 8593.08740 & 0.63708 & AC5 & 450.3 & 80.8 & 3354.6 $\pm$ \05.4 & 2739.7 $\pm$ \05.3 \\
7890.05405 & 0.53604 & MC1 & 451.0 & 73.1 & 2915.0 $\pm$  31.9 & 2223.5 $\pm$  31.2 & 8593.09386 & 0.64155 & AR1 & 601.2 & 62.0 & 3521.2 $\pm$  16.5 & 2812.0 $\pm$  15.8 \\
7890.95544 & 0.15929 & MC1 & 454.9 & 74.7 & 3363.2 $\pm$  42.0 & 2544.7 $\pm$  41.4 & 8593.10075 & 0.64631 & AG1 & 458.0 & 92.9 & 3253.2 $\pm$ \05.4 & 2604.0 $\pm$ \04.7 \\
7890.97404 & 0.17215 & MC1 & 453.7 & 74.2 & 3336.9 $\pm$  50.3 & 2709.7 $\pm$  48.3 & 8594.24926 & 0.44043 & AG1 & 457.6 & 92.9 & 2644.6 $\pm$ \04.6 & 2404.6 $\pm$ \04.8 \\
7897.02441 & 0.35558 & MC1 & 451.2 & 73.2 & 2824.5 $\pm$  29.7 & 2806.1 $\pm$  30.1 & 8594.25570 & 0.44488 & AR1 & 601.1 & 62.0 & 3028.8 $\pm$  18.3 & 2610.0 $\pm$  17.2 \\
7897.93913 & 0.98805 & MC1 & 454.7 & 74.6 & 2927.8 $\pm$  33.8 & 2286.0 $\pm$  36.5 & 8594.30703 & 0.48038 & AG1 & 457.8 & 92.9 & 2739.3 $\pm$ \04.6 & 2272.3 $\pm$ \04.9 \\
7897.95640 & 0.99999 & MC1 & 453.6 & 74.2 & 2960.0 $\pm$  33.8 & 2226.9 $\pm$  34.0 & 8594.31308 & 0.48455 & AR1 & 601.2 & 62.0 & 3121.3 $\pm$  16.5 & 2510.4 $\pm$  17.6 \\
7897.97432 & 0.01238 & MC1 & 452.8 & 73.8 & 2962.9 $\pm$  34.0 & 2165.7 $\pm$  32.7 & 8594.31964 & 0.48909 & AC5 & 449.6 & 80.7 & 2858.9 $\pm$ \05.6 & 2440.5 $\pm$ \05.4 \\
7897.99187 & 0.02452 & MC1 & 452.1 & 73.5 & 3018.8 $\pm$  33.4 & 2168.9 $\pm$  33.3 & 8595.30705 & 0.17182 & AR1 & 601.2 & 62.0 & 3435.0 $\pm$  46.4 & 2840.3 $\pm$  46.7 \\
7898.00961 & 0.03678 & MC1 & 451.6 & 73.3 & 3034.4 $\pm$  32.1 & 2172.3 $\pm$  32.9 & 8597.31539 & 0.56045 & AR1 & 601.2 & 62.0 & 3326.1 $\pm$  15.6 & 2531.9 $\pm$  25.9 \\
7898.02750 & 0.04915 & MC1 & 451.1 & 73.1 & 3111.1 $\pm$  32.4 & 2237.0 $\pm$  32.1 & 8599.06394 & 0.76946 & AC5 & 450.5 & 80.9 & 3129.0 $\pm$ \07.2 & 3048.9 $\pm$ \06.4 \\
7898.04537 & 0.06151 & MC1 & 450.7 & 73.0 & 3126.9 $\pm$  31.5 & 2250.0 $\pm$  30.9 & 8599.07006 & 0.77369 & AG1 & 458.2 & 93.0 & 2926.4 $\pm$ \05.7 & 2908.3 $\pm$ \06.8 \\
7898.06416 & 0.07450 & MC1 & 450.5 & 72.9 & 3128.2 $\pm$  28.7 & 2300.1 $\pm$  31.4 & 8599.07705 & 0.77853 & AR1 & 601.2 & 62.0 & 3335.9 $\pm$  18.2 & 3027.1 $\pm$  17.8 \\
7898.95496 & 0.69043 & MC1 & 453.6 & 74.2 & 3351.9 $\pm$  31.2 & 2620.4 $\pm$  30.7 & 8599.32386 & 0.94918 & AG1 & 458.1 & 92.9 & 2651.8 $\pm$ \05.2 & 2395.5 $\pm$ \04.4 \\
7898.97399 & 0.70359 & MC1 & 452.7 & 73.8 & 3456.7 $\pm$  30.1 & 2761.1 $\pm$  29.0 & 8599.32952 & 0.95309 & AR1 & 601.2 & 62.0 & 3096.3 $\pm$  17.3 & 2598.6 $\pm$  16.9 \\
7898.99364 & 0.71718 & MC1 & 452.0 & 73.5 & 3400.0 $\pm$  29.3 & 2730.5 $\pm$  30.1 & 8601.31700 & 0.32730 & AG1 & 458.1 & 92.9 & 2811.2 $\pm$ \04.3 & 2910.3 $\pm$ \04.5 \\
8013.90879 & 0.17340 & MC2 & 451.8 & 73.4 & 3340.5 $\pm$  31.2 & 2778.2 $\pm$  31.1 & 8602.31569 & 0.01783 & AR1 & 601.2 & 62.0 & 3123.0 $\pm$  17.9 & 2595.4 $\pm$  17.7 \\
8013.93267 & 0.18991 & MC2 & 452.8 & 73.8 & 3334.8 $\pm$  30.9 & 2871.8 $\pm$  31.6 & 8602.32331 & 0.02310 & AG1 & 458.2 & 92.9 & 2845.4 $\pm$ \05.2 & 2214.6 $\pm$ \05.5 \\
8013.95519 & 0.20548 & MC2 & 453.8 & 74.3 & 3190.9 $\pm$  31.3 & 2881.6 $\pm$  34.0 & 8605.98606 & 0.55565 & AG1 & 459.6 & 93.0 & 2950.7 $\pm$ \07.2 & 2259.4 $\pm$ \05.2 \\
8016.89170 & 0.23588 & MC2 & 451.6 & 73.3 & 3234.2 $\pm$  32.0 & 3021.4 $\pm$  32.3 & 8605.99358 & 0.56085 & AC5 & 453.4 & 81.5 & 3113.1 $\pm$  12.1 & 2383.0 $\pm$ \09.4 \\
8016.93931 & 0.26880 & MC2 & 453.4 & 74.1 & 3137.0 $\pm$  34.1 & 3029.4 $\pm$  33.1 & 8662.04050 & 0.31358 & WG1 & 458.4 & 93.0 & 2843.7 $\pm$  17.9 & 2928.8 $\pm$  15.7 \\
8016.95877 & 0.28226 & MC2 & 454.6 & 74.6 & 3157.4 $\pm$  34.0 & 3047.4 $\pm$  35.1 & 8663.95434 & 0.63688 & WG1 & 458.6 & 93.0 & 3228.6 $\pm$  17.8 & 2611.8 $\pm$  18.2 \\
8016.97699 & 0.29486 & MC2 & 456.1 & 75.2 & 3037.0 $\pm$  38.6 & 3182.2 $\pm$  37.1 & 8665.00634 & 0.36427 & WG1 & 458.4 & 93.0 & 2699.7 $\pm$  16.7 & 2812.8 $\pm$  16.7 \\
8016.99567 & 0.30777 & MC2 & 458.1 & 76.0 & 2971.9 $\pm$  40.3 & 3024.7 $\pm$  43.0 & 8669.12883 & 0.21470 & AG2 & 458.1 & 92.9 & 3099.3 $\pm$ \04.0 & 2840.8 $\pm$ \03.5 \\
8027.88105 & 0.83429 & MC2 & 452.3 & 73.6 & 2951.0 $\pm$  31.9 & 3035.6 $\pm$  32.9 & 8669.13804 & 0.22107 & AR2 & 601.2 & 62.0 & 3501.5 $\pm$  14.2 & 3105.6 $\pm$  14.2 \\
8027.90347 & 0.84979 & MC2 & 453.2 & 74.0 & 2828.7 $\pm$  31.6 & 2766.4 $\pm$  31.2 & 8669.85122 & 0.71418 & AG2 & 458.6 & 93.0 & 3233.2 $\pm$ \03.4 & 2711.4 $\pm$ \03.3 \\
8027.92614 & 0.86547 & MC2 & 454.5 & 74.5 & 2815.4 $\pm$  32.6 & 2749.4 $\pm$  32.7 & 8670.86732 & 0.41675 & AR2 & 601.2 & 62.0 & 3001.3 $\pm$  19.8 & 2857.9 $\pm$  27.5 \\
8027.94636 & 0.87945 & MC2 & 456.1 & 75.2 & 2838.9 $\pm$  32.9 & 2691.6 $\pm$  33.3 & 8670.88316 & 0.42770 & AG2 & 458.1 & 92.9 & 2578.8 $\pm$ \06.2 & 2469.1 $\pm$ \03.3 \\
8027.97027 & 0.89598 & MC2 & 458.8 & 76.2 & 2799.4 $\pm$  35.4 & 2766.7 $\pm$  34.4 & 8673.00108 & 0.89210 & AG2 & 457.5 & 92.9 & 2666.1 $\pm$ \04.2 & 2543.1 $\pm$ \03.8 \\
8039.87804 & 0.12942 & MC2 & 453.6 & 74.2 & 3365.6 $\pm$  34.7 & 2479.1 $\pm$  36.0 & 8673.16343 & 0.00436 & AG2 & 458.7 & 93.0 & 2757.3 $\pm$ \04.3 & 2221.8 $\pm$ \04.0 \\
8039.89730 & 0.14273 & MC2 & 454.8 & 74.7 & 3272.0 $\pm$  34.9 & 2690.4 $\pm$  35.8 & 8673.89816 & 0.51238 & AG2 & 457.9 & 92.9 & 2800.7 $\pm$ \03.7 & 2202.1 $\pm$ \03.6 \\
8039.91783 & 0.15693 & MC2 & 456.6 & 75.4 & 3671.7 $\pm$  41.0 & 4004.5 $\pm$  45.7 & 8673.90641 & 0.51807 & AR2 & 601.2 & 62.0 & 3240.8 $\pm$  14.1 & 2565.3 $\pm$  14.5 \\
8042.88333 & 0.20738 & MC2 & 454.4 & 74.5 & 3344.5 $\pm$  33.1 & 3267.6 $\pm$  32.9 & 8674.15167 & 0.68766 & AG2 & 458.5 & 93.0 & 3128.5 $\pm$ \03.7 & 2685.1 $\pm$ \03.6 \\
8042.90859 & 0.22484 & MC2 & 456.5 & 75.4 & 3299.5 $\pm$  32.3 & 2870.7 $\pm$  32.9 & 8674.99103 & 0.26802 & AG2 & 457.5 & 92.9 & 2909.2 $\pm$ \04.4 & 2887.2 $\pm$ \04.1 \\
8043.88464 & 0.89972 & MC2 & 454.7 & 74.6 & 2804.7 $\pm$  34.2 & 2694.5 $\pm$  34.4 & 8674.99758 & 0.27255 & AR2 & 601.1 & 62.0 & 3379.0 $\pm$  14.9 & 3127.6 $\pm$  14.8 \\
8043.90657 & 0.91488 & MC2 & 456.6 & 75.4 & 2850.8 $\pm$  34.7 & 2545.3 $\pm$  35.5 & 8676.17654 & 0.08772 & AG2 & 459.2 & 93.0 & 3004.0 $\pm$ \04.4 & 2326.6 $\pm$ \03.9 \\
8043.92603 & 0.92833 & MC2 & 458.9 & 76.3 & 2838.2 $\pm$  37.6 & 2594.7 $\pm$  36.9 & 8676.18606 & 0.09431 & AR2 & 601.4 & 61.9 & 3409.8 $\pm$  20.0 & 2694.4 $\pm$  19.4 \\
8044.87854 & 0.58693 & MC2 & 454.5 & 74.5 & 3265.1 $\pm$  34.8 & 2392.9 $\pm$  35.3 & 8677.84396 & 0.24063 & AR2 & 601.2 & 62.0 & 3430.7 $\pm$  15.7 & 3161.3 $\pm$  15.7 \\
8044.89608 & 0.59906 & MC2 & 455.9 & 75.1 & 3379.3 $\pm$  34.6 & 2488.8 $\pm$  34.8 & 8677.85367 & 0.24735 & AG2 & 458.2 & 93.0 & 3021.8 $\pm$ \04.2 & 2881.1 $\pm$ \04.2 \\
8044.91353 & 0.61112 & MC2 & 457.7 & 75.8 & 3298.1 $\pm$  34.5 & 2659.0 $\pm$  34.7 & 8678.87558 & 0.95393 & AG2 & 457.9 & 92.9 & 2641.8 $\pm$ \04.3 & 2285.4 $\pm$ \04.2 \\
8044.93233 & 0.62412 & MC2 & 460.2 & 76.8 & 3406.3 $\pm$  36.7 & 2555.1 $\pm$  37.0 & 8678.88456 & 0.96014 & AR2 & 601.2 & 62.0 & 3100.1 $\pm$  14.7 & 2609.6 $\pm$  15.7 \\
8047.90380 & 0.67869 & MC2 & 457.6 & 75.8 & 3525.7 $\pm$  36.8 & 2769.1 $\pm$  36.8 & 8704.87542 & 0.93110 & WG2 & 458.4 & 93.0 & 2584.6 $\pm$  18.5 & 2433.2 $\pm$  19.9 \\
8047.92404 & 0.69269 & MC2 & 460.2 & 76.8 & 3245.9 $\pm$  37.9 & 2950.4 $\pm$  37.7 & 8704.99069 & 0.01080 & WG2 & 458.7 & 93.0 & 2752.1 $\pm$  19.6 & 2202.1 $\pm$  19.3 \\
8047.94201 & 0.70512 & MC2 & 463.8 & 78.1 & 3230.4 $\pm$  40.5 & 2862.1 $\pm$  41.7 & 8705.07072 & 0.06613 & WG2 & 459.7 & 93.0 & 2933.6 $\pm$  18.5 & 2263.1 $\pm$  18.6 \\
8049.88856 & 0.05103 & MC2 & 456.5 & 75.3 & 3135.8 $\pm$  34.7 & 2260.4 $\pm$  34.7 & 8705.94277 & 0.66910 & WG2 & 458.5 & 93.0 & 3193.8 $\pm$  17.9 & 2663.6 $\pm$  18.4 \\
8049.90678 & 0.06363 & MC2 & 458.6 & 76.2 & 2978.2 $\pm$  35.5 & 2427.1 $\pm$  35.2 & 8706.91827 & 0.34359 & WG2 & 458.4 & 93.0 & 2740.5 $\pm$  16.3 & 2798.7 $\pm$  15.5 \\
8049.92508 & 0.07628 & MC2 & 461.4 & 77.3 & 3276.8 $\pm$  38.5 & 2337.4 $\pm$  37.7 & 8707.00354 & 0.40255 & WG2 & 458.8 & 93.0 & 2498.4 $\pm$  18.9 & 2611.0 $\pm$  18.4 \\
8153.26473 & 0.52881 & AC1 & 454.4 & 67.1 & 3297.0 $\pm$ \05.4 & 2616.9 $\pm$ \05.0 & 8718.06573 & 0.05133 & WG3 & 460.5 & 93.0 & 2906.1 $\pm$  19.1 & 2198.2 $\pm$  19.1 \\
8153.27290 & 0.53445 & AC1 & 454.0 & 67.0 & 3325.0 $\pm$ \06.8 & 2603.3 $\pm$ \06.8 & 8718.90832 & 0.63392 & WG3 & 458.5 & 93.0 & 3242.1 $\pm$  17.4 & 2538.2 $\pm$  17.1 \\
8154.26944 & 0.22350 & AC2 & 454.0 & 67.0 & 3638.6 $\pm$ \06.8 & 3375.4 $\pm$ \06.3 & 8718.99914 & 0.69672 & WG3 & 459.2 & 93.0 & 3211.8 $\pm$  16.8 & 2798.6 $\pm$  17.6 \\
8211.31589 & 0.66735 & AC3 & 457.6 & 69.7 & 3524.1 $\pm$ \06.2 & 2990.7 $\pm$ \06.6 & 8719.06301 & 0.74088 & WG3 & 460.5 & 93.0 & 3038.9 $\pm$  17.2 & 2855.6 $\pm$  18.0 \\
8212.31267 & 0.35655 & AC3 & 457.6 & 69.7 & 2892.7 $\pm$ \06.1 & 3113.1 $\pm$ \06.2 & 8719.99780 & 0.38722 & WG3 & 459.2 & 93.0 & 2664.9 $\pm$  17.1 & 2674.5 $\pm$  16.4 \\
8213.30765 & 0.04452 & AC3 & 457.6 & 69.6 & 3214.4 $\pm$ \06.6 & 2554.0 $\pm$ \06.7 & 8759.91091 & 0.98449 & WG4 & 459.6 & 93.0 & 2744.0 $\pm$  20.2 & 2313.8 $\pm$  19.7 \\
8214.30728 & 0.73570 & AC3 & 457.6 & 69.7 & 3183.6 $\pm$ \06.6 & 3077.8 $\pm$ \06.0 & 8779.89481 & 0.80205 & WG5 & 460.7 & 93.0 & 2775.5 $\pm$  19.7 & 2959.4 $\pm$  19.9 \\
8215.31262 & 0.43082 & AC3 & 457.7 & 69.7 & 2781.3 $\pm$ \07.4 & 2565.3 $\pm$ \06.3 & 8779.92552 & 0.82328 & WG5 & 462.1 & 93.0 & 2757.6 $\pm$  22.0 & 2914.0 $\pm$  22.5 \\
8216.31191 & 0.12176 & AC3 & 457.7 & 69.7 & 3431.3 $\pm$ \06.8 & 2597.3 $\pm$ \06.3 & 8780.91512 & 0.50752 & WG5 & 461.7 & 93.0 & 2774.9 $\pm$  22.2 & 2339.6 $\pm$  23.9 \\
8343.97692 & 0.39367 & MC3 & 451.0 & 73.1 & 2794.3 $\pm$  44.1 & 2699.4 $\pm$  44.7 & 8782.90202 & 0.88133 & WG5 & 461.3 & 93.0 & 2596.1 $\pm$  20.7 & 2786.5 $\pm$  19.7 \\
8343.99015 & 0.40281 & MC3 & 451.3 & 73.2 & 2844.1 $\pm$  47.0 & 2642.1 $\pm$  45.6 & 8782.93426 & 0.90362 & WG5 & 463.5 & 93.1 & 2587.2 $\pm$  24.3 & 2617.8 $\pm$  25.1 \\
\hline
\end{tabular}
\end{table*}

Between April 2017 and October 2019, 147 high precision polarimetric observations were made of the close binary system $\mu^1$~Sco (HD~151890) in three different pass bands. A smaller number of observations of its wide companion $\mu^2$~Sco (HD~151985) were also made, along with observations of low and high polarization standards for calibration purposes. These observations were made using three different telescopes and two different HIPPI-class polarimeters in varying combinations. Mini-HIPPI \citep{Bailey17} was used on a 35-cm Celestron C14 telescope at UNSW Observatory, which is located on campus in an inner suburb of Sydney, Australia. HIPPI-2 \citep{Bailey20} was used both on the 3.9-m Anglo Australian Telescope (AAT) located at Siding Spring Observatory, and on the 60-cm Ritchey-Chretien telescope at the Western Sydney University (WSU) Penrith Observatory. Table \ref{tab:runs} gives a summary of the set-up for each observing run for each filter used.

HIPPI-2 and Mini-HIPPI were both used in the study of the Spica binary system \citep{Bailey19}. HIPPI-2 has also recently been used in the study of the rapidly rotating system $\alpha$~Oph \citep{Bailey20b}, the red supergiant Betelgeuse \citep{Cotton20a} and the polluted white dwarf G29-38 \citep{Cotton20b}. 

All the observations used Hamamatsu H10720-210 photo-multiplier tubes for detectors. The $\mu^1$~Sco observations were either made without a filter (Clear) or using Astrodon SDSS $g^{\prime}$ or $r^{\prime}$ filters. Used without a filter Mini-HIPPI and HIPPI-2 have sensitivity between $\sim$350 and 700~nm, resulting in a similar effective wavelength to the SDSS $g^{\prime}$ filter. The formal error for each observation is the square root of the sum of the squares of the internal measurement precision and a positioning error. The positioning error, which is different in different filter bands and on different telescopes is described in \citet{Bailey20}. In order to beat seeing noise, HIPPI-class polarimeters use Ferroelectric Liquid Crystals (FLCs) to modulate the sign of polarization at a frequency of 500~Hz. Three different FLCs were utilised as outlined in Table \ref{tab:runs}; their performance characteristics are described in \citet{Bailey20}. Standard observing procedures and reduction, as described in \citet{Bailey15, Bailey17, Bailey20} were used, except that we are now calculating airmass to two decimal places, for improved precision on higher polarization objects. We have also improved the bandpass model for the Celestron C14 by including the transmission of the corrector plate, which is modelled as 5.49~mm MgF$_{\rm2}$ coated iron float glass. For the purposes of the bandpass model $\mu^1$~Sco is modelled as spectral type B1 and $\mu^2$~Sco as B2. For most observations a single sky (S) measurement was made adjacent to each target (T) measurement at each of the four position angles, $PA=0, 45, 90, 135^{\circ}$, in the pattern TSSTTSST. During twilight when necessary the first or last target measurement was sometimes bracketed between two skies to account for the rapid changes in sky polarization.

A small polarization due to the telescope mirrors, TP, shifts the zero-point offset of our observations. This is corrected for by reference to the straight mean of several observations of low polarization standard stars, details of which are given either in \citet{Bailey20} or in the caption of Table \ref{tab:observations}. Similarly, the position angle (PA) is calibrated by reference to literature measurements of high polarization standards, also given in either \citet{Bailey17}, \citet{Bailey20} or in the caption of Table \ref{tab:runs}. TP calibrations are made in the same band as the observations, while the PA calibration is initially made with observations in SDSS $g^{\prime}$ and Clear, with corrections applied for other bands based on a smaller number of observations. Correction of a minor software glitch that sometimes induced a 0.3$^{\circ}$ PA error, has improved the PA precision in later runs for HIPPI-2 over what was reported in \citet{Bailey20}.

Fig. \ref{fig:observations} displays the $\mu^1$~Sco observations, which are also listed in Table \ref{tab:observations}. The tabulated phases have been computed based on the ephemeris of \citet{vanAntwerpen10} (see also Section \ref{sec:mu1Sco_intro}). 

The right hand panel of Fig. \ref{fig:observations} shows the data plotted on a Q-U diagram. An approximately elliptical locus of points is traced out here in each of the three filter bands, which is the shape expected from an intermediately inclined system. The three loops are offset from each other, as one would expect if the zero points are constant but different, as in the case of them being set by interstellar polarization. The $r^{\prime}$ data clearly has a tighter loop (i.e. a lower polarization amplitude) than $g^{\prime}$ or Clear data. This is what we expect for a reflection mechanism \citep[][Supplementary Information]{Bailey19}. Compared to Spica, the peak-to-trough amplitude is about four times larger. Given the comparative temperatures of the components of the binaries and the shorter orbital period of the $\mu^1$~Sco system, this is also consistent with photospheric reflection.

\subsection{Statistical analysis}
\label{sec:stats}

\begin{table*}
\centering
\caption{Moment calculations.}
\tabcolsep 3.5 pt
\begin{tabular}{lc|cccccc|ccccccc}
\hline
\hline
Filter & n      & \multicolumn{6}{c|}{$q$} & \multicolumn{6}{c}{$u$}\\
    &           & Mean & Mean Err. & Std. Dev. & Err. Var. & Kurtosis & Skewness & Mean & Mean Err. & Std. Dev. & Err. Var. & Kurtosis & Skewness\\
\hline
Clear (MH)      & 70 & 3041.5 & 34.6\phantom{$\mathsection$} & 237.5 & 235.0  & 0.164\phantom{**} & 2.223\phantom{**} & 2608.0 & 34.6\phantom{$\mathsection$} & 335.2 & 333.4 & 1.055** & 5.354** \\
Clear (H2)      & 24 & 3278.4 & 10.2\phantom{$\mathsection$} & 222.9 & 222.7  & 0.592\phantom{**} & 2.614\phantom{**} & 2765.2 & \phantom{0}9.7\phantom{$\mathsection$} & 215.7 & 215.5 & 0.017\phantom{**} & 1.868\phantom{**} \\
$g^{\prime}$    & 38 & 2864.5 & 11.7$\mathsection$ & 213.4 & 213.1  & 0.180\phantom{**} & 2.072\phantom{**} & 2571.7 & 11.5$\mathsection$ & 258.1 & 257.9 & 0.002\phantom{**} & 1.514** \\
$r^{\prime}$    & 16 & 3284.0 & 18.6\phantom{$\mathsection$} & 167.4 & 166.4  & 0.069\phantom{**} & 1.645*\phantom{*} & 2822.4 & 19.6\phantom{$\mathsection$} & 331.3 & 230.4 & 0.009\phantom{**} & 1.418** \\
\hline
Clear (MH)\dag  & 69 & 3035.2 & 34.5\phantom{$\mathsection$} & 226.9 & 224.2  & 0.082\phantom{**} & 1.908**           & 2596.9 & 34.5\phantom{$\mathsection$} & 292.7 & 290.7 & 0.052\phantom{**} & 2.060*\phantom{*} \\
Clear (H2)\ddag & 12 & 3115.3 & \phantom{0}6.5\phantom{$\mathsection$} & 197.3 & 197.2 & 0.046\phantom{**} & 1.891\phantom{**} & 2766.7 & \phantom{0}6.0\phantom{$\mathsection$} & 277.3 & 277.2 & 0.012\phantom{**} & 1.287** \\
\hline
\hline
\end{tabular}
\label{tab:stats}
\begin{flushleft}
Notes: \dag~Data point at JD$=$2458039.9178 removed. \ddag~AC3 and AC5 only. $\mathsection$~$g^{\prime}$ data includes both AAT and WSU data, the mean errors of which in $q$~/~$u$ are respectively 4.7~/~4.3~ppm and 18.9~/~19.0~ppm. MH: Mini-HIPPI, H2: HIPPI-2. Error variance (Err. Var.) is $\sqrt{(x^2-e^2)}$, where $x$ is the standard deviation (Std. Dev.) and $e$ the mean error (Mean Err.) of a set of measurements; all of these quantites, along with the means are in ppm. Skewness is defined so that 3 is the normal value. The asterisks indicate significance in skewness or kurtosis (calculated using the tables of \citealp{Brooks94}), one asterisk for 95 per cent, two for 99 per cent. \\
\end{flushleft}
\end{table*}

We examined the variability of our polarimetric data set with a moment analysis in \ref{tab:stats}, as is often done before looking at phases \citep{Clarke10, Brooks94}. The data has been divided up by filter. In the case of the Clear observations we also split the data by instrument, since the bandpass with \mbox{HIPPI-2} varied a bit depending on the modulator condition and whether or not the Barlow lens had to be used. It is important to note here that while differences in the modulation efficiency curve for the modulators do not affect the effective wavelength, they do result in different weightings for different wavelengths to the measured polarization; this is important for the later eras of the BNS performance (see \citealp{Bailey20}). 

The Error Variance is a measure of polarimetric variability not accounted for by known uncertainties; in Table \ref{tab:stats} it has values between $\sim$165 and $\sim$335~ppm. Which confirms what is already obvious in Fig. \ref{fig:observations} -- that there is intrinsic variability in the system. 

There is also a significant negative skewness in the $u$ Stokes parameter in the $g^{\prime}$ and $r^{\prime}$ data (and less significantly in $q$). Which means that the tail of the distribution is on the left of the Q-U plot. Skewness can indicate an intermediate inclination of the orbit in a binary system, with its sign dependant on the position angle of the line of nodes. 

At first glance it does not appear the Clear data has the same characteristics as $g^{\prime}$ and $r^{\prime}$ in Table \ref{tab:stats}, but interrogating the data more closely reveals it is consistent.  The AC4 run observations were all made in quick succession, and have similar phase (Table \ref{tab:observations}); these corresponded to a period of heavy cloud when observing other targets wasn't possible. They also taken during the BNS modulator's Era~7, which has the most extreme drift in $\lambda_{\rm 0}$. When we remove this data, and the three points from AC1 and AC2 during HIPPI-2's commissioning, where modifications to the back-end of the instrument were being made on the fly, and just use data from AC3 and AC5 the anomaly disappears. Similarly there is an obvious outlier in the Mini-HIPPI data at JD$=$2458039.9178 (see Fig. \ref{fig:observations}), if we remove it this data is also consistent.

\subsection{\texorpdfstring{$\mu^2$ Sco}{mu2~Sco}}

In order to obtain an estimate to the interstellar polarization of $\mu^1$~Sco independent of our modelling, we also made measurements of its wide companion $\mu^2$ Sco. Two Mini-HIPPI observations were obtained during the MC1 run, and a HIPPI-2 observation each in $g^{\prime}$ and $r^{\prime}$ in AG1 and AR1 respectively. A HIPPI-2 observation in Clear was made during run AC5. The same calibrations in TP and PA were applied as were for $\mu^1$~Sco, and the other details of the observations are the same too. The results of the observations are shown in Table \ref{tab:mu2Sco_obs}. The tabulated results probably under-estimate the errors for the Mini-HIPPI observations, perhaps by a factor of 2, or 3 at most. $\mu^2$~Sco is not as bright as the low polarization objects used to estimate the errors on the C14 at UNSW, and variability in sky conditions will be more significant.

The polarizations measured for $\mu^2$~Sco are similar to (but not the same as) the mean values given for $\mu^1$~Sco in Table \ref{tab:stats}, suggesting that the majority of the constant polarization is interstellar in origin. Notably the values are higher in $r^{\prime}$ than in $g^{\prime}$ or Clear, indicating a redder value of $\lambda_{max}$ -- the wavelength of maximum polarization -- consistent with stars in the wall of the Local Hot Bubble \citep{Cotton19b}, or beyond it \citep{Serkowski75}, which have values around 550~nm.

\begin{table}
\caption{High precision polarimetric observations of $\mu^2$ Sco.}
\label{tab:mu2Sco_obs}
\tabcolsep 5.25pt
\centering
\begin{tabular}{cccccc}
 \hline
 HJD & Run    &   $\lambda_{eff}$ & Eff & $q$ & $u$ \\
 2450000+ && (nm) & (\%) & (ppm) & (ppm) \\
\hline
7872.04108 & MC1 & 454.4 & 74.6 & 3009.8 $\pm$ 37.0 & 2526.3 $\pm$ 36.0 \\
7884.03179 & MC1 & 453.6 & 74.2 & 3169.3 $\pm$ 37.0 & 2413.9 $\pm$ 36.7 \\
8596.21739 & AC5 & 450.0 & 80.9 & 3158.0 $\pm$ \phantom{0}5.3 & 2540.5 $\pm$ \phantom{0}5.4 \\
8600.31973 & AR1 & 601.2 & 62.0 & 3452.4 $\pm$ 20.6 & 2796.4 $\pm$ 22.9 \\
8601.32407 & AG1 & 458.5 & 92.9 & 3028.2 $\pm$ \phantom{0}6.5 & 2461.6 $\pm$ \phantom{0}6.4 \\
\hline
\end{tabular}
\end{table}

\subsection{Comparison with previous observations}
\label{sec:comparison}

For easy comparison the observations of \citet{Serkowski75} and \citet{Smith56} are given in terms of $q$ and $u$ in Table \ref{tab:previous}. They are probably best compared to the Mini-HIPPI observations. The observations of $\mu^2$~Sco are in agreement within error, but this is unremarkable given the size of the errors in the old measurements. \citet{Serkowski70}'s observations of $\mu^1$~Sco are of the same scale as those reported here, but the agreement is at best fair when one considers the phases.

\begin{table}
\caption{Previous polarimetric observations of $\mu^1$~Sco and $\mu^2$~Sco.}
\label{tab:previous}
\setlength\extrarowheight{2pt}
\tabcolsep 3.25pt
\centering
\begin{tabular}{cccccc}
\hline
\hline
Reference & Star    &   $\lambda_{eff}$ & $q$   & $u$   & Phase \\
    &         &   (nm)            & (ppm) & (ppm) &       \\
\hline
\citet{Serkowski70}  & $\mu^1$~Sco & $\sim$435 & 2701~$\pm$~\phantom{0}100 & 2431~$\pm$~\phantom{0}100 & 0.616\\ 
\citet{Serkowski70}  & $\mu^1$~Sco & $\sim$435 & 3312~$\pm$~\phantom{0}100 & 2406~$\pm$~\phantom{0}100 & 0.108 \\
\hline
\citet{Smith56} & $\mu^2$~Sco & $\sim$450 & 2960~$\pm$~1302 & 3527~$\pm$~1302 & \\
\hline
\hline
\end{tabular}
\end{table}

\section{Modelling}
\label{sec:modelling}

The approach to modelling the polarization of a binary system follows the methods described by \citet{Bailey19}. We model only the polarization effects arising from the photospheres of the two stars, including reflection of light from one star off the other and vice versa, tidal distortion of the stars, and the eclipse effect \citep{Chandrasekhar46,Kemp83}. 

These methods are an extension of those we have used for modelling the polarization of rapidly rotating stars \citep{Cotton17, Bailey20b}. We map the distribution of temperature and gravity over the surface of the stars and interpolate in a set of stellar atmosphere models to determine the spectral radiance (specific intensity) and polarization of the light emitted and reflected from each point on the stellar surface. These values can then be integrated over the stars to give the total light and polarization as a function of orbital phase.

\subsection{Binary geometry}
\label{sec:geom}

The treatment of binary geometry follows the methods of \citet{Wilson71} and \citet{Wilson79}. The surfaces of the stars are modelled as equipotential surfaces in a Roche model using an extended definition of the potential $\Omega$ as given in equation 1 of \citet{Wilson79}. We use a cartesian coordinate system in which the centre of the primary star is at the origin, the line joining the two stars is along the $x$ axis and the binary orbit is in the $xy$ plane. Coordinates are expressed in units of the orbital semi-major axis. The methods allow for eccentric orbits and non-synchronous rotation as was the case in our modelling of Spica \citep{Bailey19}. However, in the case of $\mu^1$ Sco we consider only the simpler case of a circular orbit and synchronous rotation. 

At a given point on the stellar surface with coordinates $x, y, z$, the local effective gravity is given by the vector
\begin{equation}
\mathbfit{g}_{\rm eff} = \begin{bmatrix} \partial\Omega/\partial x, \partial\Omega/\partial y, \partial\Omega/\partial z \end{bmatrix}.
\end{equation}
The local normal to the surface \mathbfit{n} is in the opposite direction and thus given by
\begin{equation}
\mathbfit{n} = -\mathbfit{g}_{\rm eff}/\lvert\mathbfit{g}_{\rm eff}\rvert.
\end{equation}
If $\mathbfit{o}$ is the normalized vector to the observer (the direction of which will depend on orbital inclination and orbital phase), then
\begin{equation}
    \mu = \mathbfit{n} \cdot \mathbfit{o}
\end{equation}
where $\mu$ is the cosine of the local viewing zenith angle. If $\mu < 0$ then the point is on the unseen side of the star.

Our treatment of reflected light requires the direction to the source. In our geometry this vector $\mathbfit{s}$ is:
\begin{equation}
\mathbfit{s} = \begin{bmatrix} D-x, -y, -z \end{bmatrix}.
\end{equation}
where the source is taken to be the centre of the other star at a distance $D$ along the x-axis, and for a circular orbit, $D = 1$ (with an eccentric orbit D would vary with orbital phase). Then
\begin{equation}
    \mu_0 = \mathbfit{n} \cdot (\mathbfit{s}/\lvert\mathbfit{s}\rvert)
\end{equation}
where $\mu_0$ is the cosine of the illuminating zenith angle. If $\mu_0 < 0$ then the point is not illuminated by the source. The angle between the plane containing $\mathbfit{o}$ and $\mathbfit{n}$ and that containing $\mathbfit{s}$ and $\mathbfit{n}$ is the azimuthal angle $\phi-\phi_0$. The three geometrical parameters $\mu$, $\mu_0$ and $\phi-\phi_0$ are needed as inputs for the radiative transfer calculation.  

The above discussion applies to points on the surface of the primary star. The secondary star is treated in the same way by moving the coordinate system origin to the centre of the secondary and defining an internal potential according to equation 2 of \citet{Wilson79}.

\subsection{Gravity darkening}

We assume that gravity and effective temperature are related by the von Zeipel law, $T_{\rm eff} \propto g^\beta$ with $\beta = 0.25$ \citep{vonzeipel24}. A recent model of gravity darkening in binary stars by \citet{Espinosa12} is based on the assumption that energy flux is anti-parallel to effective gravity. It shows deviations from the von Zeipel law, particularly at extreme mass ratios, but for binaries where the two stars have similar masses the effective $\beta$ remains close to the von Zeipel value of 0.25. Using this relation and the Roche model we can then determine the $T_{\rm eff}$ at any point on the star's surface given the polar value.

\subsection{Stellar atmosphere models}
\label{sec:atmos}

For each star we calculate a set of \textsc{atlas9} stellar atmosphere models covering a range of $\log{g}$ and $T_{\rm eff}$ values appropriate for each star. The models are based on the solar composition model grids of \cite{Castelli03}. Because of our assumption of the von Zeipel relationship, a set of models derived by varying one parameter ($\log{g}$) are sufficient to cover each star.

\begin{figure*}
    \centering
    \includegraphics[width=14cm] {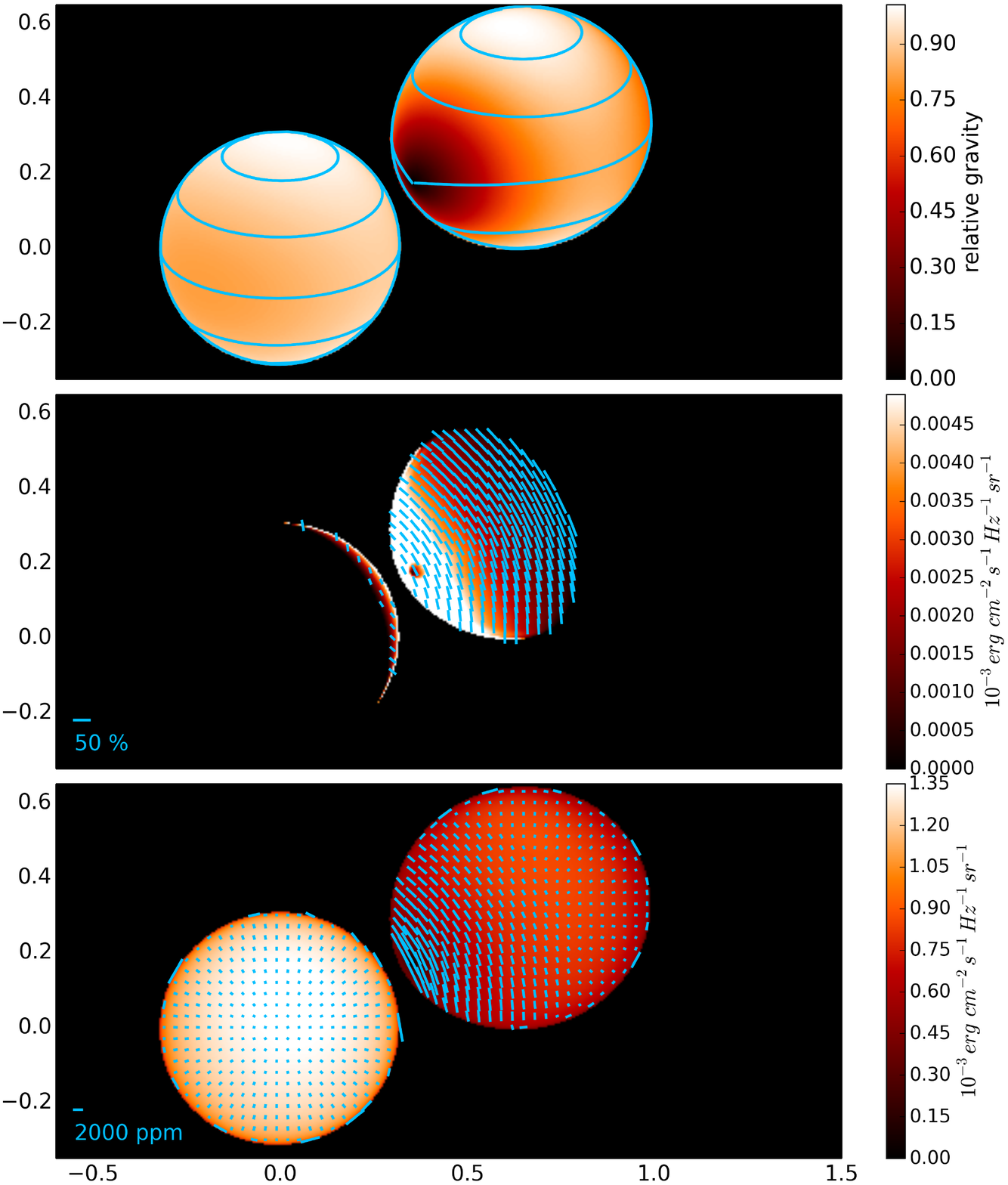}
    \caption{Example of polarization modelling of $\mu^1$ Sco at a phase of 140$\degr$ (=0.3$\dot{8}$). The upper panel shows the gravity distribution of each star relative to the polar value. The middle panel shows the spectral radiance (specific intensity) overlaid with polarization vectors for the reflected light only. The lower panel is the same but including both emitted and reflected light. Intensity is plotted for every pixel, polarization vectors are plotted for every sixth pixel in each direction. These model results are for a wavelength of 450 nm.}
    \label{fig:mod_image}
\end{figure*}

\subsection{Polarized radiative transfer}
\label{sec:radtran}

To calculate the spectral radiance (specific intensity) and polarization of the radiation from these stellar atmosphere models, we use a version of the \textsc{synspec} stellar spectral synthesis code \citep{hubeny85,hubeny12}, modified to include polarized radiative transfer using the \textsc{vlidort} code of \citet{spurr06}. \textsc{vlidort} (Vector Linearized Discrete Ordinate Radiative Transfer) is an implementation of the discrete ordinate method of radiative transfer, incorporating a full treatment of polarization. It has been used in Earth atmosphere research as well as planetary atmosphere modelling \citep{Bailey18,Bott18}. In \citet{Cotton17} and \citet{Bailey20b} \textsc{synspec/vlidort} was used for modelling the polarization of a rotationally distorted star, and the supplementary materials of the first work included verifications of the methods by comparison with stellar polarization calculations of \citet{Harrington15} and rotating star polarization models of \citet{Sonneborn82}.

The polarization effects included in the analysis are Thomson scattering from electrons as well as Rayleigh scattering from H, He and H$_2$ all of which are described by a Rayleigh scattering matrix. Electron scattering is the main effect at the temperatures of the $\mu^1$ Sco components.

We do separate radiative transfer calculations for the emitted light and reflected light from the star. For the emitted light we calculate the intensity and polarization from each stellar atmosphere model as a function of wavelength and viewing angle ($\mu$). The wavelength grid is non-uniform and is chosen by \textsc{synspec} to fully sample the line structure of the spectrum. We re-bin to a uniform wavelength spacing of 0.04-nm for further analysis.

For reflected light the methods are similar, but there are now three geometric parameters, the viewing zenith angle ($\mu$), the illuminating zenith angle ($\mu_0$), and the azimuthal angle between the two ($\phi - \phi_0$) as described in section \ref{sec:geom}. We calculate intensity and polarization for each wavelength and for a coarse grid of geometric parameters (9 $\mu$ values, 9 $\mu_0$ values, 13 azimuths) and use 3D spline interpolation in this grid to obtain values for any required geometry.

\subsection{Integration over the stars}

For a given inclination and orbital phase we set up a rectangular grid of pixels covering the observer's view of the star and with a grid spacing of 0.005 times the orbital semi-major axis. For each pixel that overlaps one of the stars, and is not occulted by the other star, we determine the local effective gravity (and hence the effective temperature) and the geometric parameters as described in section \ref{sec:geom}. We then interpolate in the set of stellar atmosphere radiative transfer results as described in sections \ref{sec:atmos} and \ref{sec:radtran}. We interpolate first in gravity and then in geometry to obtain an intensity and polarization value for each pixel. These values can be plotted on an image as in Fig. \ref{fig:mod_image}, or the values can be summed over the two stars to give the integrated polarization as a function of phase.

\begin{table}
    \tabcolsep 2.5 pt
    \centering
    \caption{Parameters of $\mu^1$~Sco system used as polarization model input.}
    \begin{tabular}{lll}
    \hline
    Parameter    &  Value   &   Comments  \\
    \hline
    \multicolumn{3}{c}{Adopted parameters from \citet{vanAntwerpen10}}\\
    \hline
    Primary Mass ($M_1$) & 8.49 $\pm$ 0.05 \msun  &  \\
    Primary $T_{\rm eff}$ ($T_1$) & 23725 $\pm$ 500 K &  (mean temp.) \\
    Primary Potential ($\Omega_1$) &  3.85 $\pm$ 0.01 &    \\
    Secondary Mass ($M_2$) & 5.33 $\pm$ 0.05 \msun & \\
    Secondary $T_{\rm eff}$ ($T_2$) & 16850 $\pm$ 500 K &  (mean temp.) \\
    Mass ratio ($q_b$) & 0.627 $\pm$ 0.004 & \\
    Inclination ($i$) & 65.4 $\pm$ 1.0$\degr$ & \\
    Eccentricity ($e$) & 1.0 & \\
    Semi-major axis ($a$) & 12.90 $\pm$ 0.04 \rsun & \\
    \hline
    \multicolumn{3}{c}{Additional input parameters for polarization model (derived from above)}\\
    \hline
    Primary Polar Temp. ($T1_p$) & 24350 K  \\
    Primary Polar Gravity ($\log{g1_p}$) & \multicolumn{2}{l}{4.169 \phantom{000001} (from $M_1$, $\Omega_1$, $q_b$, $a$)}\\
    Secondary Polar Temp. ($T2_p$) & 18110 K  \\
    Secondary Polar Gravity ($\log{g2_p}$) & 3.939 & (from $M_2$. $q_b$, $a$) \\
    \hline \hline
    \end{tabular}
    \label{tab:modparam}
\end{table}

\subsection{Polarization models of \texorpdfstring{$\mu^1$ Sco}{mu1 Sco}}
\label{sec:models}

Our modelling of $\mu^1$ Sco is based on the parameters for the system determined by \citet{vanAntwerpen10} from fitting to the light curve and radial velocity data. The parameters used as input to our modelling are listed in Table \ref{tab:modparam}. We note that an alternate analysis of the system by \citet{Budding15} gives very similar parameters. The relative geometry of the binary system is determined by the mass ratio, here denoted $q_b$, and the potentials of the two stars. The primary potential is as listed in Table \ref{tab:modparam} and the secondary potential is constrained by the requirement that it fills its Roche lobe (i.e. a semi-detached binary as also assumed by \citealt{vanAntwerpen10}). The rotation of both stars is assumed to be synchronized with the orbital period.

Our modelling code requires the polar temperature and gravity of each star as input values. The temperatures listed by \citet{vanAntwerpen10} are mean temperatures, the polar temperatures were calculated from these allowing for the tidal distortion of the stars and assuming the \citet{vonzeipel24} gravity darkening law. The polar gravity is calculated from the masses listed in Table \ref{tab:modparam}, the relative radii and the orbital semi-major axis as listed in the table.

\section{Results and Analysis}
\label{sec:results}

\subsection{Modelling Results} 
\label{sec:modres}

\begin{figure}
    \centering
    \includegraphics[width=\columnwidth, trim={0.25cm 0.8cm 0.5cm 0.8cm},clip]{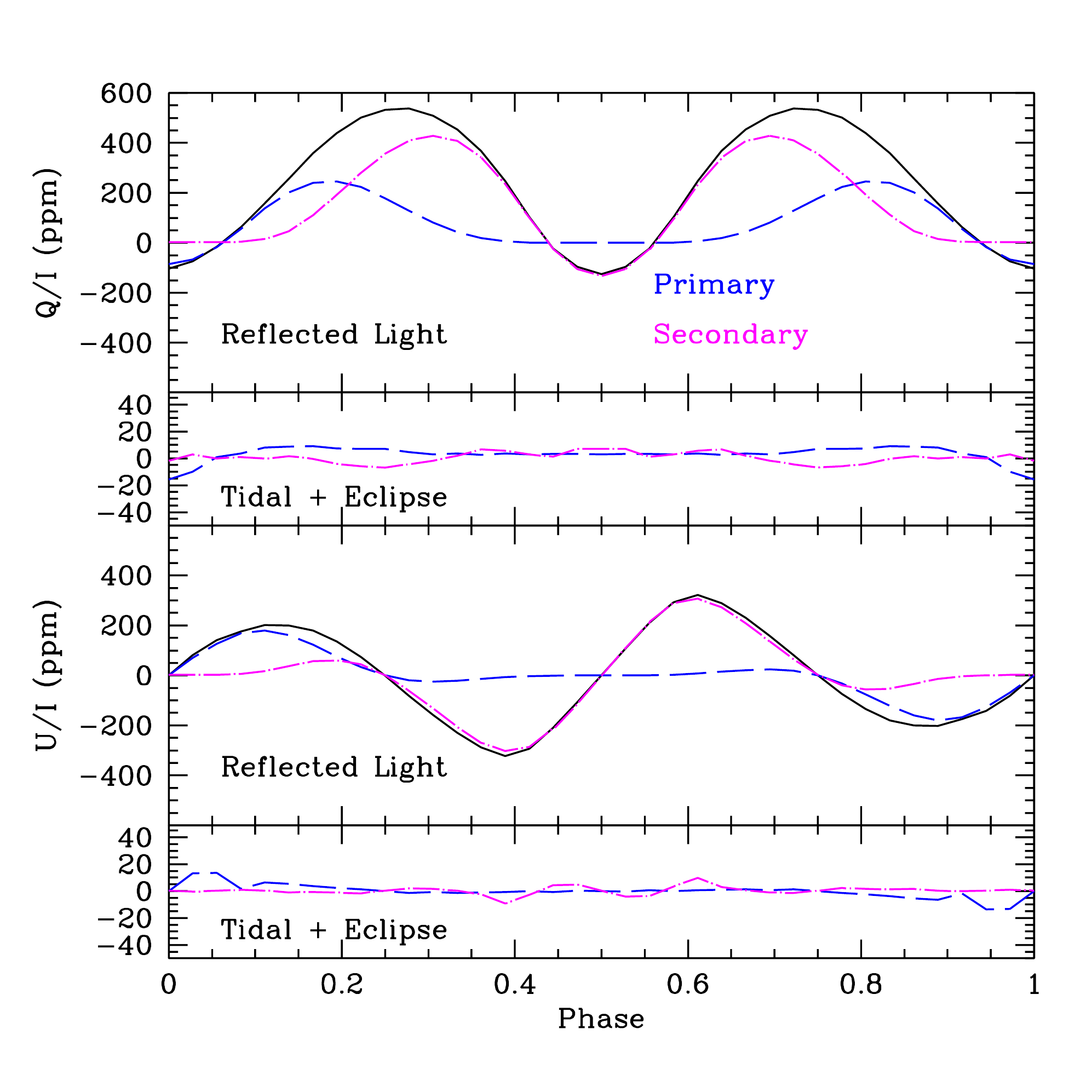}
    \caption{Components making up the total phase dependent modelled polarization of $\mu^1$ Sco at 450~nm. Contributions from the primary component (dashed blue lines), and from the secondary (dot-dash magenta lines) are split into the contribution of reflected light, and the thermal emission which includes the tidal and eclipse effects. All contributions are shown as a fraction of the total light from both components. The black lines show the total polarization as used in the fits in Fig. \ref{fig:phase} (a -- c).}
    \label{fig:comps}
\end{figure}

To compare our observational data with the models we have calculated them at 35 phase angles at two wavelengths -- 450~nm and 600~nm -- since the calculations are computationally very expensive. The fitted model curves include contributions from reflection, tidal distortion and an eclipse effect, all for both primary and secondary components. The break-down of these components is shown for the 450~nm case in Fig. \ref{fig:comps}. 

Two features of Fig. \ref{fig:comps} are worth noting. Firstly, reflection is the most important effect, with tidal forces making a much smaller contribution. The magnitude of the eclipse effect is larger than the the tidal forces, but still relatively small. Compared to Spica \citep{Bailey19}, whose components are more massive but wider separated, the tidal effect is proportionally smaller in $\mu^1$~Sco. Secondly, the contributions of both components are important, and consequently the curves have a more complicated shape than in modelled hot-Jupiter exoplanet systems \citep{Seager00, Bott18}.

In the 600~nm case (not shown) the phase curves look similar. However, the amplitude of the polarization from reflection is smaller -- about two thirds that of the 450~nm case. Conversely, the eclipse effect is about fifty percent larger for the 600~nm calculation.

\begin{figure*}
\centering
\subcaptionbox{Mini-HIPPI Clear data, 450~nm fit.}{\includegraphics[width=.475\textwidth]{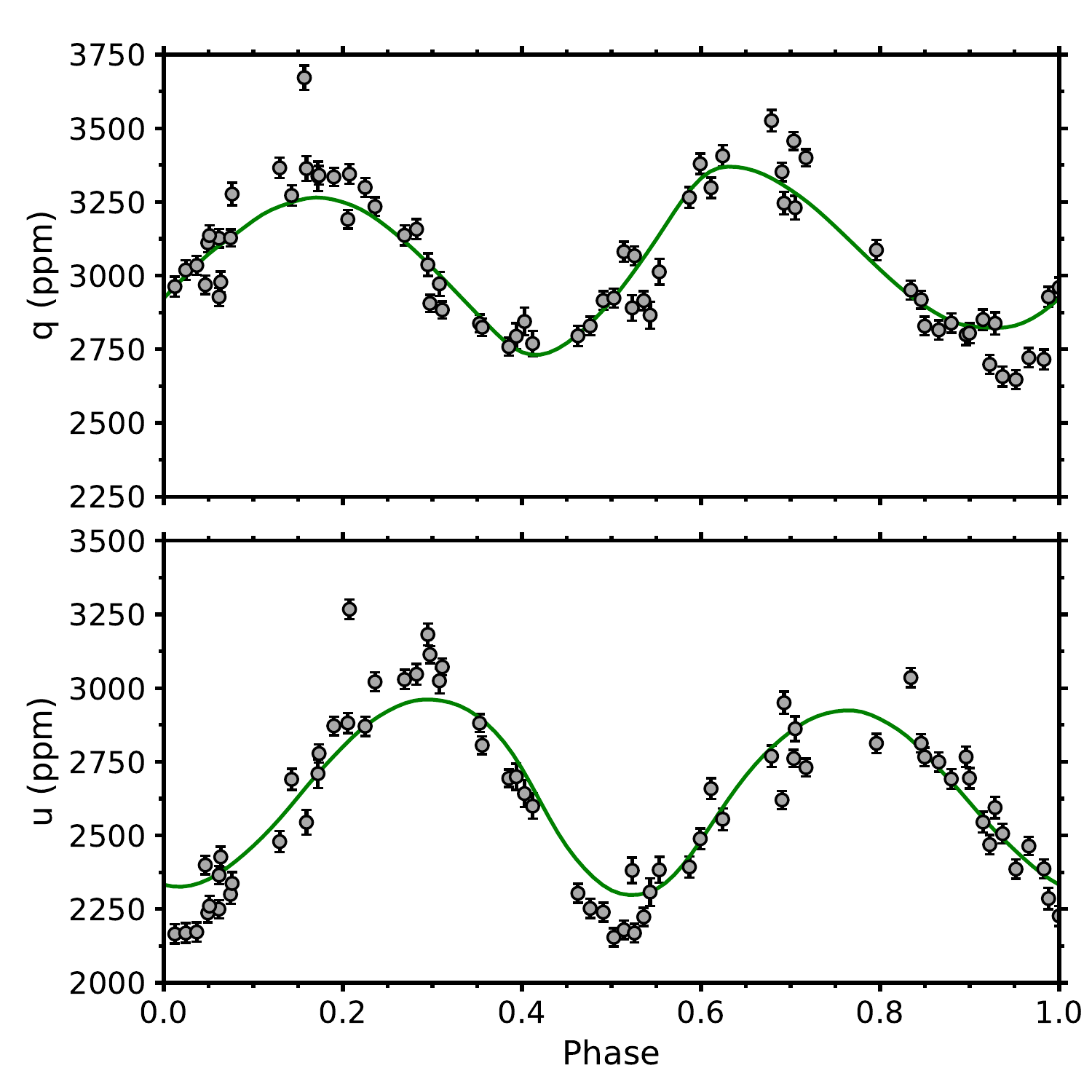}\quad}
\subcaptionbox{HIPPI-2 Clear data, 450~nm fit.}{\includegraphics[width=.475\textwidth]{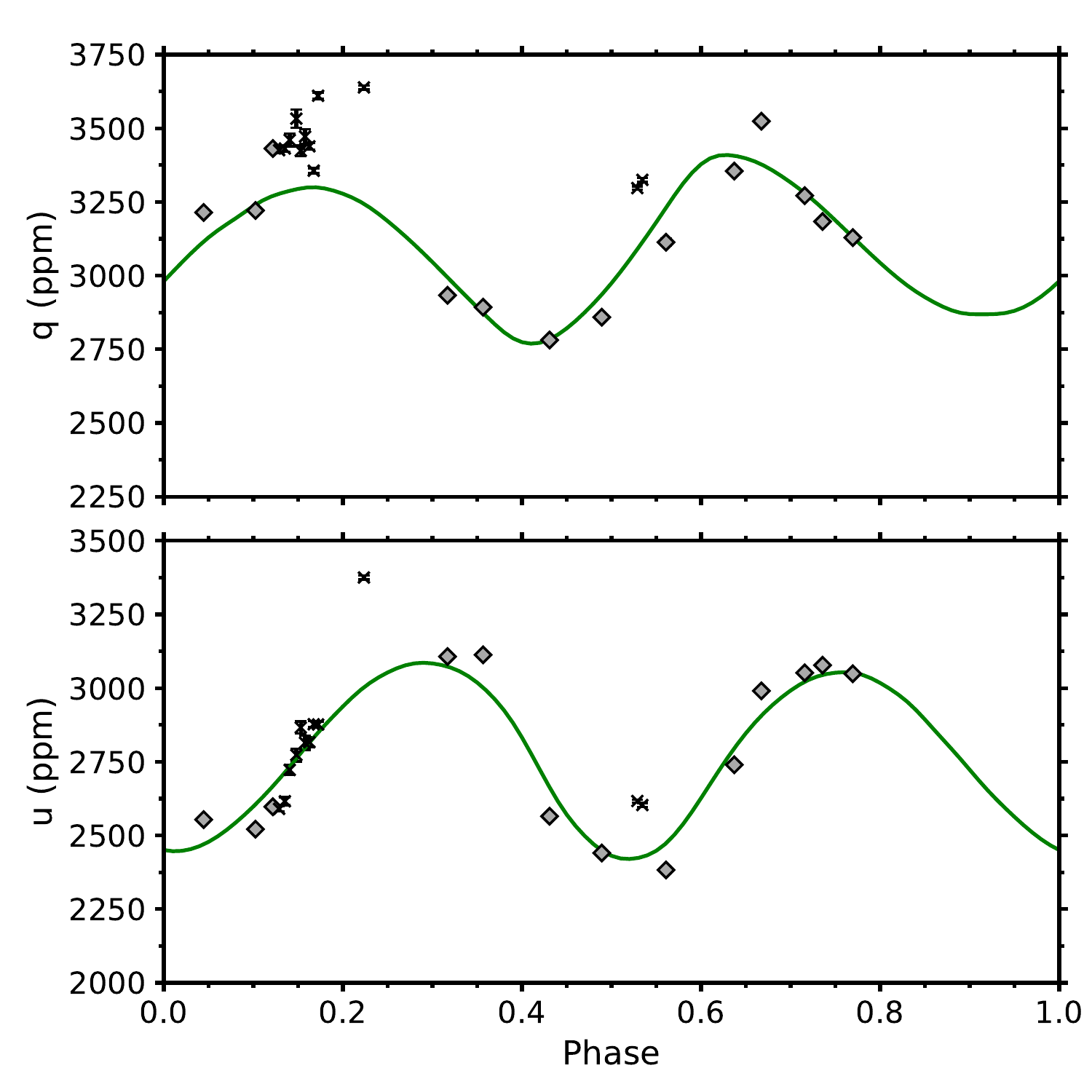}\quad}
\subcaptionbox{HIPPI-2 $g^{\prime}$ data, 450~nm fit.}{ \includegraphics[width=.475\textwidth]{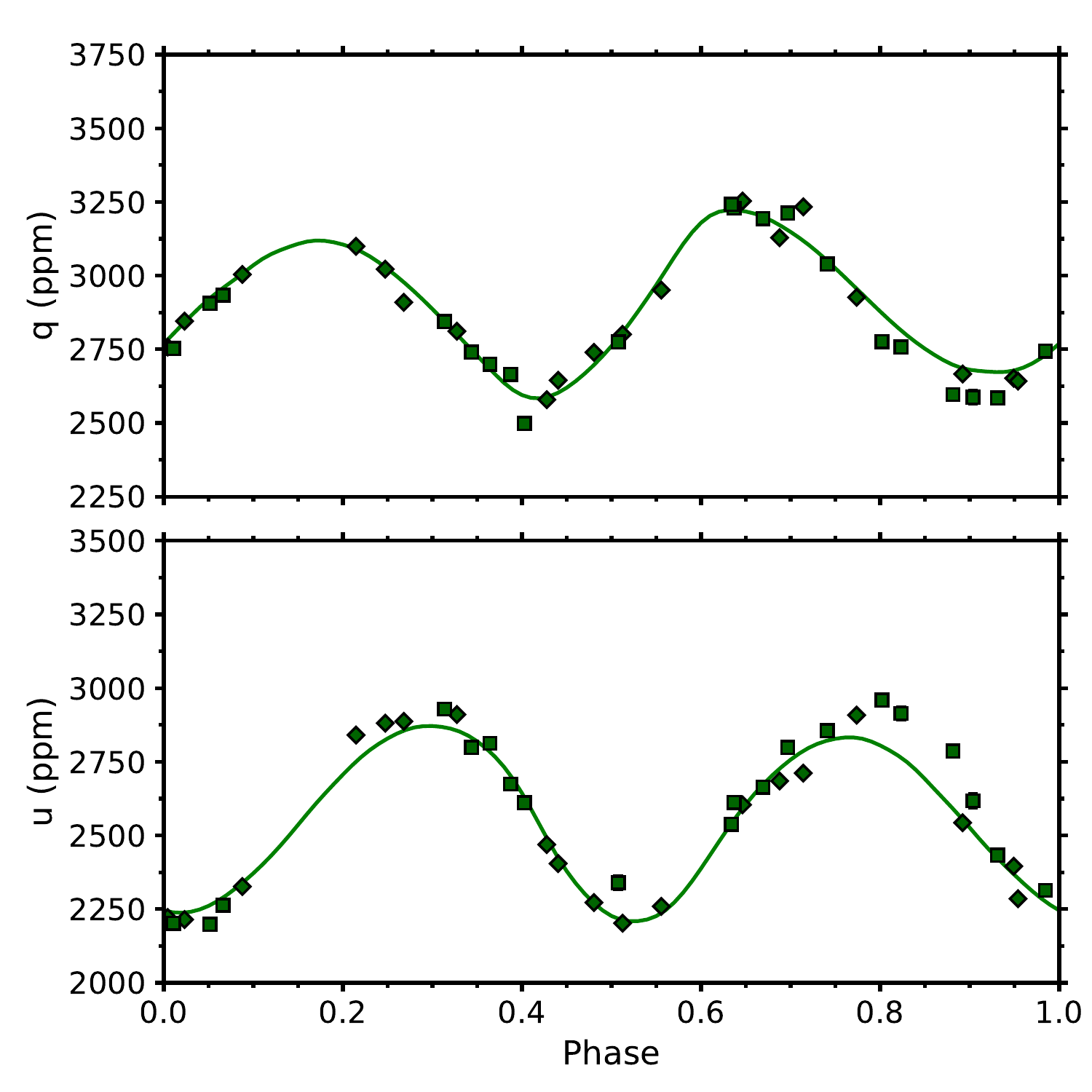}\quad}
\subcaptionbox{HIPPI-2 $r^{\prime}$ data, 600~nm fit.}{ \includegraphics[width=.475\textwidth]{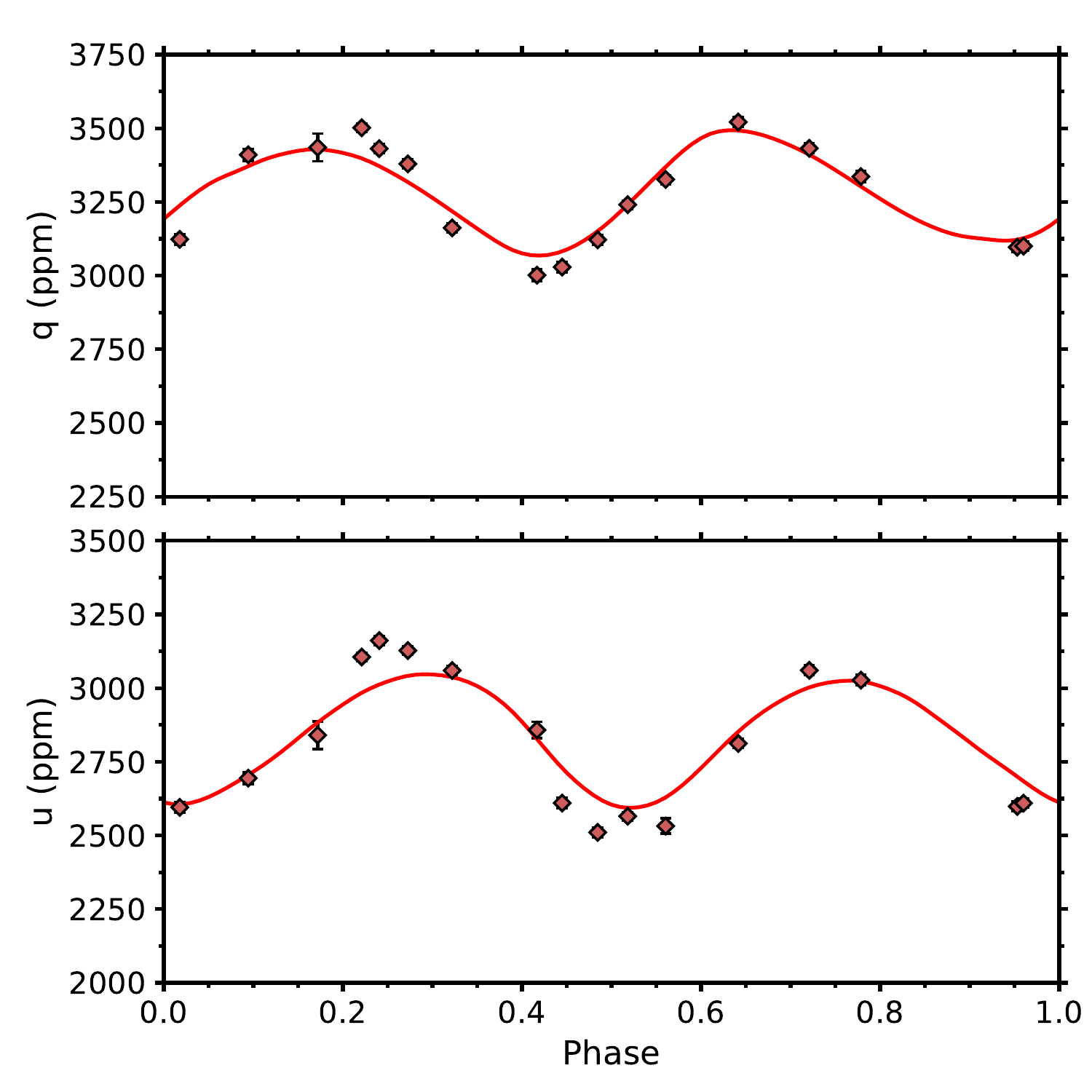}}
\hfill
\caption{Polarization phase curves showing monochromatic fits (450~nm -- green lines; 600~nm -- red line) to different observational data. Clear observations are in grey, SDSS $g^{\prime}$ in dark green, SDSS $r^{\prime}$ in light red, UNSW data are circles, WSU squares and AAT diamonds (runs AC3 and AC5) and crosses (runs AC1, AC2 and AC4 not used for the fit. Fit curves are calculated for phase intervals of 0.02$\dot{\rm 7}$ (10$^\circ$) and interpolated in between.}
\label{fig:phase}
\end{figure*}

\subsection{Model Fitting}
\label{sec:modfit}

To find the best fit we use a Levenberg-Marquardt nonlinear least-squares algorithm \citep{Press92} with just three parameters: the offsets in each of the linear polarization Stokes parameters, $Z_q$ and $Z_u$, and the position angle of the line of nodes, $\Omega$ -- there are no parameters that adjust the amplitude of the polarization.  The results of our modelling fits to various data sets are shown in Fig. \ref{fig:phase}, and listed in Table \ref{tab:fit_pars} are the fitted parameters and their uncertainties (determined using a bootstrap analysis based on 1000 random samplings from the available observations).

\begin{table}
\caption{Fit parameters.}
\label{tab:fit_pars}
\tabcolsep 8 pt
\begin{tabular}{lccc}
\hline
\hline
Filter & $Z_q$ (ppm)& $Z_u$ (ppm)& $\Omega$ ($^\circ$)\\
\hline
Clear (MH)   & 2962.6 $\pm$ 13.2 & 2428.1 $\pm$ 15.2 & 123.84 $\pm$ 1.02 \\
Clear (H2)\ddag   & 3014.5 $\pm$ 30.4 & 2548.7 $\pm$ 18.1 & 125.60 $\pm$ 1.77 \\
$g^{\prime}$ & 2810.6 $\pm$ \07.4 & 2340.7 $\pm$ \09.5 & 123.21 $\pm$ 1.17 \\
$r^{\prime}$ & 3222.3 $\pm$ 13.8 & 2686.3 $\pm$ 24.4 & 124.06 $\pm$ 1.38 \\
\hline
\hline
\end{tabular}
\begin{flushleft}
Notes: MH: Mini-HIPPI, H2: HIPPI-2. \ddag AC3 \& AC5 only.
\end{flushleft}
\end{table}

The values of $Z_q$ and $Z_u$ are similar to the mean values of $q$ and $u$ given in Table \ref{tab:stats}, typically differing by around 100~ppm in any given band. This means they are also similar to the measured values for $\mu^2$~Sco (Table \ref{tab:mu2Sco_obs}), and thus represent close to the interstellar polarization values. Assuming this to be the case, the $g^\prime$ to $r^\prime$ $Z_p$ ratio corresponds to a $\lambda_{\rm max}$ for the ISM of $\sim$650~nm -- a little redder than the typical value of 550~nm. That the $\mu^2$~Sco values do not agree exactly with the offset values may suggest differences in dust content on the sight lines of the two stars -- and as the interstellar polarization is significant and they're a fairly wide binary this is possible. Alternatively it could indicate $\mu^2$~Sco is intrinsically polarized. Most B-type stars are intrinsically polarized \citep{Cotton16}, so this is also possible.

The polarization amplitude of the 600~nm model is smaller than that of the 450~nm model. There are no fit parameters that change the polarization amplitude to fit the data. Despite this, the amplitude and shape of the polarization phase curves in Fig. \ref{fig:phase} are all very well matched by our monochromatic fits. The match to the complicated shape of the phase curves described by Fig. \ref{fig:comps} is especially evident in $q$ in the $g^\prime$ data (Fig. \ref{fig:phase}(c)). Any effect neglected by the model must fit within the much smaller magnitude described by the residuals -- this includes scattering by gas streams. However, we believe that the largest variances can be explained by the simplifications made in the models or observational inaccuracies not captured by the formal errors; in particular difficulties around $PA$ calibration.

\subsubsection{Discrepancies due to systematic observational effects}

\begin{figure}
\includegraphics[width=\columnwidth]{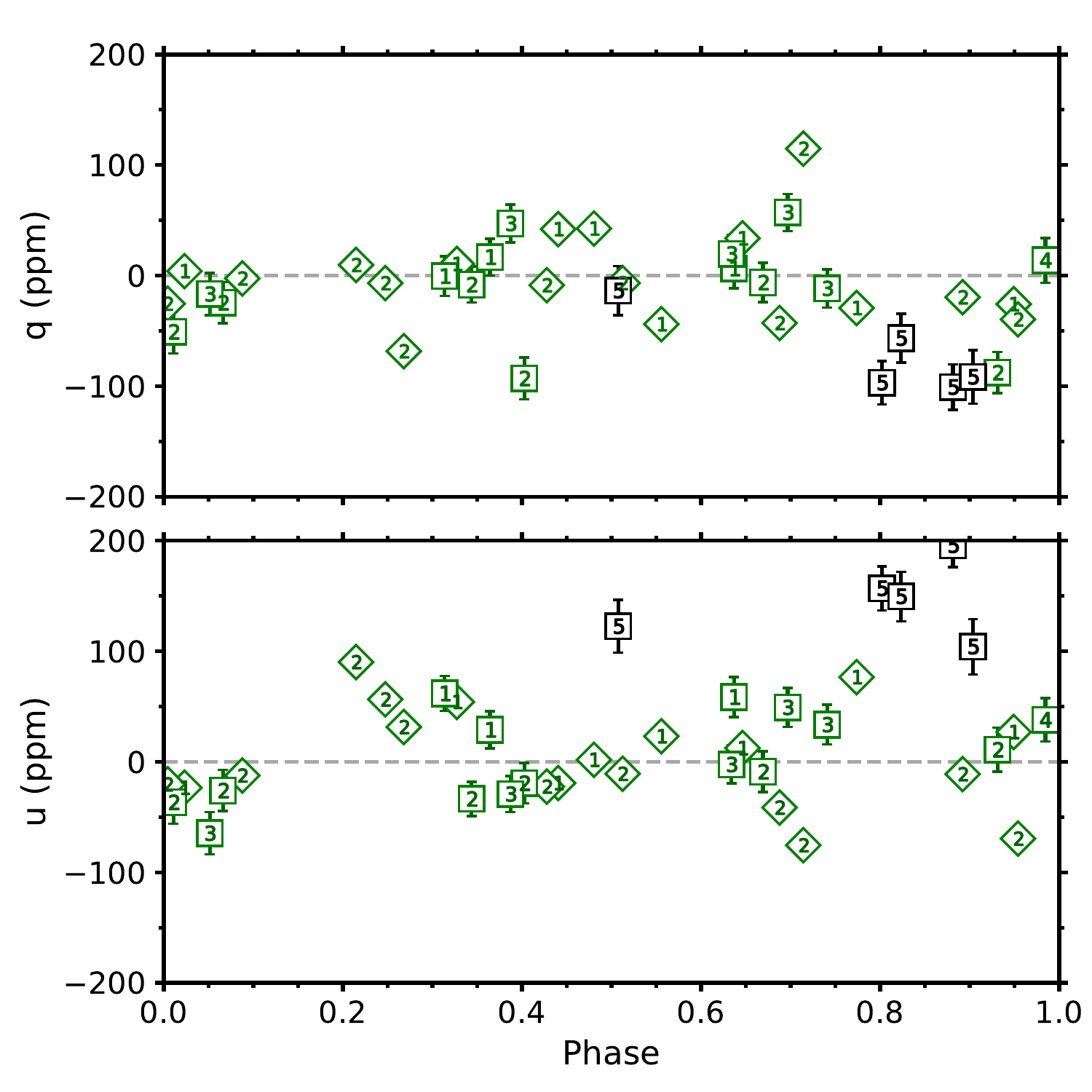}
\caption{Residuals (data $-$ model) from the fits to data presented in Fig. \ref{fig:phase}(c). Numbers and symbols indicate the telescope the data was taken with: AAT (diamonds) and WSU (squares), and the numbers the specific run. Highlighted is run WG5 - black squares annotated `5'.}
\label{fig:g_residual}
\end{figure}

An instructive example is Fig. \ref{fig:phase}(c), which shows a region around phase 0.8 to 0.9 in $u$ where the $g^\prime$ observations deviate from the model. In Fig. \ref{fig:g_residual} the residuals in this data are broken down by telescope and run, with WG5 highlighted. Run WG5 makes up most of the discrepant data at phase 0.8 to 0.9. It can be seen that where the model underestimates the data in $u$ it overestimates the data in $q$. This suggests a difference in the position angle calibration ($PA$). A simple calculation shows that a 1~degree change in position angle corresponds to $\sim$100~ppm being transferred from $q$ to $u$ at these polarization levels. 

The high polarization calibration stars we use have uncertainties of around a degree in $PA$ \citep{Bailey20}. We use a short list of standards with the aim of collecting enough comparative data to improve standard precision at a later date, but this is beyond the scope of the current paper. Here, however, we can demonstrate that the $PA$ calibration is the likely cause of the observed discrepancy with Table \ref{tab:wsu_pa_cal}. Shown in the table are the absolute $PA$ correction for each WSU run, and the relative differences to that for each standard observation, separated by standard. Every run except WG5 used an observation of HD~147084, most also used observations of HD~187929 which comparably has a positive offset. However, WG5 includes only observations of HD~187929 and HD~160529 which shows a large positive offset relative to HD~187929. The net effect of this is that WG5 data is probably rotated by nearly a degree relative to the other runs, accounting for the discrepancy to the model. The multiple observations of HD~147084 during WG1, demonstrate that the differences between calibrator stars are significant. AAT runs typically include more observations of a greater number of standards, and these effects tend to average out, but smaller offsets will also be present between runs.

\begin{table}
\caption{WSU $PA$ Calibration in $g^\prime$.}
\label{tab:wsu_pa_cal}
\tabcolsep 7 pt
\begin{tabular}{lrrrrr}
\hline
\hline
Run         & WG1\phantom{*}     & WG2     & WG3     & WG4     & WG5\phantom{\dag} \\
$PA$ Corr.  & 50.12\phantom{*}   & 95.58   & 95.45   & \05.91  & \05.89\phantom{\dag} \\
\hline
\textit{Standard} & \multicolumn{5}{c}{\textit{Relative difference to $PA$ correction}} \\
HD~147084   & $-$0.24 & $-$0.05 & $-$0.18 & $-$0.44 &         \\
            & $+$0.01 &         &         &         &         \\
            & $-$0.01 &         &         &         &         \\
            & $-$0.03 &         &         &         &         \\
            & $-$0.05 &         &         &         &         \\
            & $+$0.10 &         &         &         &         \\
            & $-$0.00 &         &         &         &         \\
            & $-$0.05 &         &         &         &         \\
            & $+$0.09 &         &         &         &         \\
            & $+$0.11 &         &         &         &         \\
HD~154445   & $+$0.08 &         &         &         &         \\
HD~160529   &         &         &         &         & $+$0.84 \\
HD~187929   &         & $+$0.05 & $+$0.18 & $+$0.44 & $-$0.25 \\
            &         &         &         &         & $-$0.59 \\
\hline
\hline
\end{tabular}
\begin{flushleft}
Notes: All values are in degrees. The typical error in an individual observation is 0.05$^{\circ}$ for the WG runs.\\
\end{flushleft}
\end{table}

A similar issue exists for the Mini-HIPPI data presented in Fig. \ref{fig:phase}(a), where 27 out of 34 observations from run MC1 fall short of the model prediction in $u$. The $PA$ calibration for MC1 includes two observations of HD~84810 which show approximately a $-$0.5$^{\circ}$ offset to HD~147084. Whereas MC2 and MC3 runs include standards showing a positive offset to HD~147084 (HD~187929, and HD~149757 which is $\sim$+1$^{\circ}$. Because of the uneven sampling, the MC1 points correspond mostly to the minimums in the $u$ curve (phases near 0.0 and 0.5). As a work-around for a wiring issue the MC1 run used different instrument PAs ($-$90$^{\circ}$, -45$^{\circ}$, 0$^{\circ}$, 45$^{\circ}$) to those used for later runs. The PA correction for MC1 is very close to 90$^{\circ}$, so any inaccuracy caused by using non-standard instrument angles may be limited to one Stokes parameter.

Most of the calibration issues for HIPPI-2 on its commissioning runs have already described in section \ref{sec:stats}. In addition to this, during early HIPPI-2 runs a minor software glitch resulted in a sporadic rotation error of $\sim$0.3$^{\circ}$ which amounts to $\sim$30~ppm being transferred between $q$ and $u$. The issue was prevalent in runs AC1--4.

Given the city-based aspect of the WSU and UNSW observatories, the sky background changes more quickly close to the horizon, and is probably not as well corrected for by a sky measurement made sequentially with the target measurement. The scatter in the Mini-HIPPI measurements is greater than for other bands. It should be noted that $\mu^1$~Sco is 1 -- 2 magnitudes fainter than the stars used to estimate the precision of the instrument on the UNSW telescope \citep{Bailey20}, and the UNSW observatory is much more badly effected by light pollution than WSU. The target is also much more difficult to centre in the aperture at UNSW because of severe backlash in the C14's mount, particularly poor centering is the most likely cause of the two most discrepant individual points in the Mini-HIPPI data.

\subsubsection{Discrepancies due to broadband effects}

\begin{figure}
\includegraphics[width=\columnwidth, trim={0.2cm 1.0cm 0.0cm 0.7cm},clip]{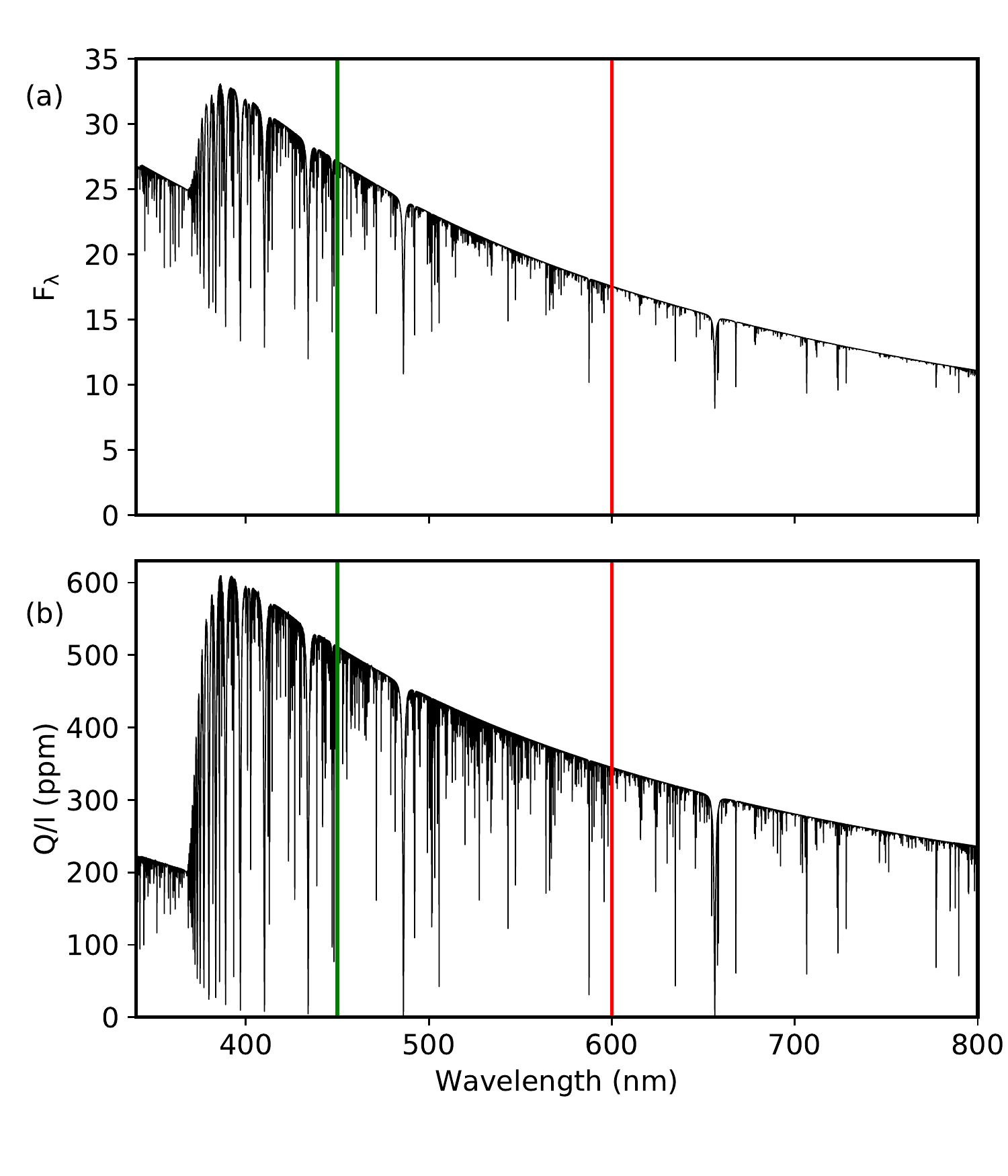}
\caption{The flux specturm (a) and polarization spectrum in $Q/I$ (b) at a phase of 110$^\circ$ (0.3056) for the $\mu^1$~Sco system (black). The green (450~nm) and red (600~nm) lines show the wavelengths of the monochromatic models.}
\label{fig:pol_spec}
\end{figure}

Systematic errors also arise as a result of fitting monochromatic models to broadband observations. The efficiency corrections applied by the bandpass model do not account for polarization changing as a function of wavelength, since this would require predicting the polarization before the observations are processed. However, once a model has been devised this can be fed into the bandpass model to see the effect. It is not practical to generate a polarization spectrum for the entire set of phases, but in Fig. \ref{fig:pol_spec}(b) the polarization spectrum for $Q/I$ at a phase of 110$^\circ$ (0.3056) is presented as an example. The polarization has a complex non-linear wavelength dependence.

Calculating the polarization using the polarization spectrum in Fig. \ref{fig:pol_spec}(b) and the detailed flux spectrum for the system in \ref{fig:pol_spec}(a) in our bandpass model reveals that the 450~nm model over-estimates the polarization by $\sim$25~ppm in both the $g^\prime$ and Mini-HIPPI passband at airmass 1.0 -- the situation is improved at higher airmasses since the lower polarization region shortward of 400~nm is attenuated -- which also accounts for some of the scatter in that data, though this is less than 10~ppm. A smaller effect due to the polarization spectrum changing with the changing contributions of the binary components will also be present correlated with phase (see Fig. \ref{fig:comps}), since the spectral types are not the same.

Together the above described observational complications will account for the majority of the scatter in the data. This cannot be due to scattering from gas, as that is a grey process and not all bands are equally affected.

Smaller residuals in the $g^\prime$ and $r^\prime$ data, that might be phase dependent, do not map as easily to the phase curves. If real these probably relate to the effects of heating on the phase-locked stars which is not currently accounted for by the model. Taking account of this is the subject of future work.

\subsection{Orbital Orientation}

\begin{figure}
    \centering
    \includegraphics[width=\columnwidth, trim={0.75cm 1.0cm 1.25cm 1.5cm}, clip]{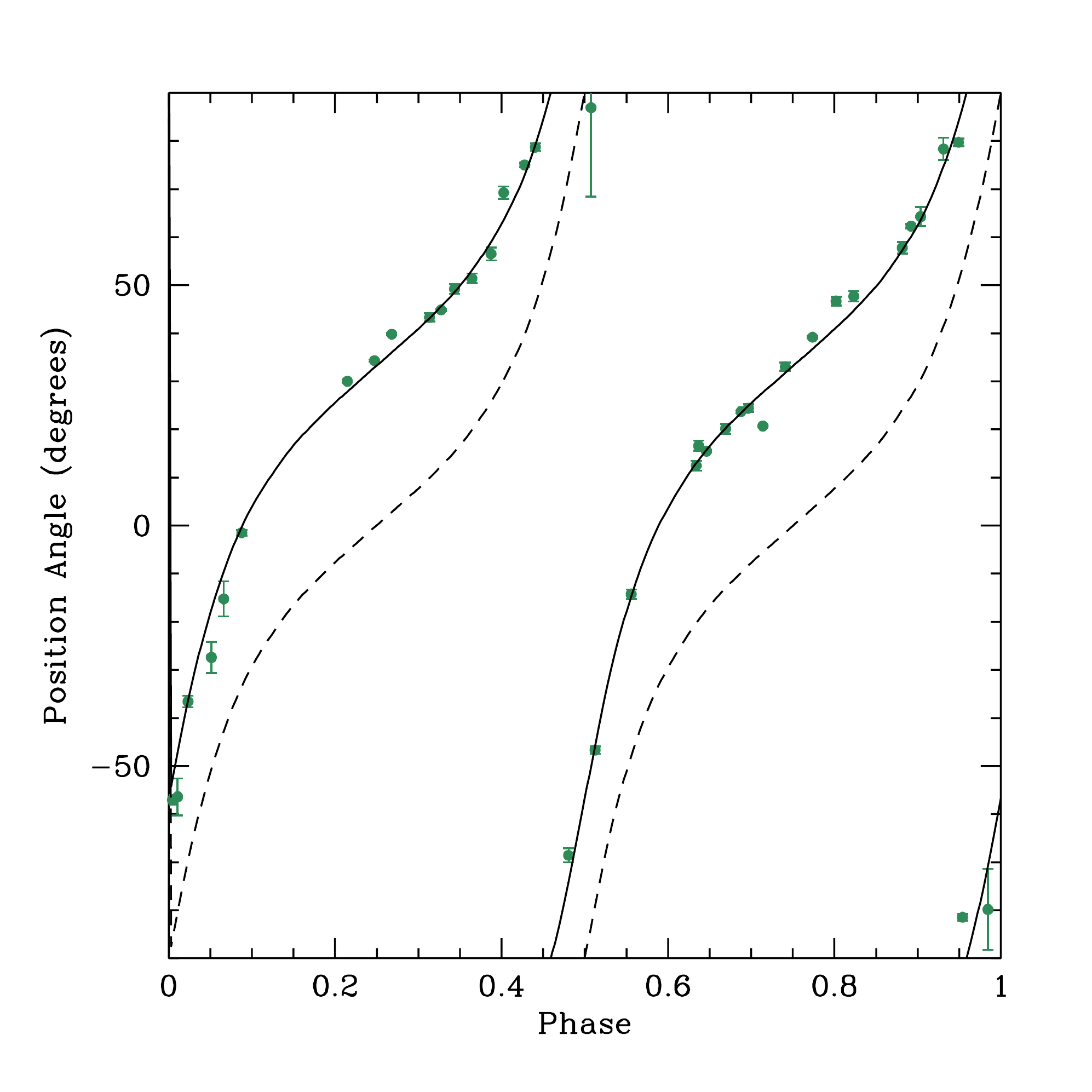}
    \caption{Polarization position angle variation shown by the $g^\prime$ band data (green points) of $\mu^1$ Sco after correcting for the polarization offsets listed in Table \ref{tab:fit_pars}. The dashed line shows the modelled polarization position angle in the model coordinate system. Rotating the model through a further 33.21$\degr$ gives the solid line which corresponds to the position angle of the line of nodes of 123.21$\degr$ as given in Table \ref{tab:fit_pars}.}
    \label{fig:pa}
\end{figure}

The fit of the models to the data determines the position angle of the line of nodes ($\Omega$) of the orbit as listed in Table \ref{tab:fit_pars}. This is further illustrated in Fig. \ref{fig:pa}. Here the polarization position angle of the $g^\prime$ observations, after correction for the offsets in Table \ref{tab:fit_pars}, is compared with the modelled position angle. The dashed line shows the position angle in the model coordinate system (e.g. as used in Fig. \ref{fig:mod_image}) which has the orbit oriented along the x-axis and corresponds to $\Omega$ = 90$\degr$. To fit the data the model needs to be rotated by a further 33.21$\degr$ giving $\Omega$ = 123.21$\degr$.

It can be seen from Fig, \ref{fig:pa} that the position angle increases with time. Since position angle is measured anticlockwise from north, the orbital direction is therefore also anticlockwise on the sky.

\section{Discussion}
\label{sec:discussion}

\subsection{The polarigenic mechanism}

The observations of $\mu^1$ Sco clearly show a double peaked polarization variation of the same form as seen in many other binary systems as discussed in section \ref{sec:intro}. In the past the most common interpretation of such polarization has been that it is due to electron scattering from optically thin circumstellar gas as described, for example, by the \textit{\mbox{BMcLE} model} \citep{Brown78}.

It has been previously shown using simplified mechanisms \citep{Berdyugin99, Berdyugin16} that photospheric reflection can match the observed double-peaked structure. Moreover, our polarized radiative transfer analysis clearly shows that the observed polarization amplitude can be entirely explained by effects due to the stars' photospheres, with reflection of light being the most important effect. Indeed the model predicts the correct amplitude for the polarization using only the input parameters derived from the light curve and radial velocity model of \citet{vanAntwerpen10}. There are no adjustable fit parameters that change the polarization amplitude. The only parameters that were fitted to the data were the polarization offsets (probably due to interstellar polarization) and the orientation of the orbit.

As well as the correct prediction of the observed polarization amplitude in both $\mu^1$ Sco and Spica \citep{Bailey19}, a number of other factors support the reflection interpretation. Firstly the reflection model correctly predicts the polarization curve shapes that have slightly different shapes for the two maxima and the two minima with no further assumptions. Secondly the reflection model predicts the correct wavelength dependence of the polarization amplitude, with smaller polarization amplitudes in the $r^\prime$ than in the $g^\prime$ band. The \textit{\mbox{BMcLE} model} with optically thin electron scattering gas predicts a wavelength independent polarization amplitude\footnote{\citet{Berdyugin16} explain a wavelength dependence in HD~48099 by a dilution effect of unpolarized free-free emission from the surrounding circumstellar matter.}. Thirdly the reflection model correctly predicts that polarization will be seen in both semi-detached binaries such as $\mu^1$ Sco and in detached binaries such as Spica, whereas on the \textit{\mbox{BMcLE} model} we would expect polarization to occur primarily in semi-detached binaries where there is mass transfer to provide a source of circumstellar gas. 

It is possible that reflection accounts for a lot of the polarization observed in early-type binaries. All hot stars have electrons in their atmosphere which lead to a small (a few per cent) of incident light being reflected, and as this light is predominantly single scattered it will be very highly ($\sim$90 \%) polarized at optimal scattering angles (see Fig. 2 of \citealt{Bailey19}). The percentage of light reflected increases with increasing temperature and decreasing surface gravity. This matches the trend of increasingly large intrinsic polarizations found in more luminous early-type binary systems \citep{Koch89}\footnote{An extreme example is the 8000~ppm amplitude seen in the O8.5I~+~O7IIIf binary HD~149404 \citep{Luna88}.}, which implies \citep{Schaffer15} photspheric reflection is a more prominent mechanism than has been credited.

This is not to imply that scattering by entrained gas is not present as a polarization mechanism in many binary systems. There is ample evidence for the presence of gas in many systems. For instance, in semidetached systems like Algol, U~Sge and CX~Dra, Doppler tomography has revealed the presence of gas streams and disks \citep{Richards00, Richards04, Richards07}. Furthermore \citet{Piirola80} showed in the U~Cep system the amplitude of polarization correlated with the changeable mass-transfer rate. Polarimetry is also an important tool in the study of Wolf-Rayet systems where it is used to investigate the nature and geometry of the stellar wind \citep{StLouis93, Johnson19, Fullard20}; undoubtedly Wolf-Rayet winds are a source of polarization.

Yet, this does not rule out photospheric reflection being an important component of the total polarization in these systems. Every ordinary star in a binary will reflect light from its photosphere, and that light will be polarized, it is only the magnitude of the effect that will vary. The results presented here and in \citet{Bailey19} suggest this could be significant. Even in WR+O systems reflection from the hot luminous photosphere of the \textit{companion} is likely to be a significant source of polarization not presently accounted for. This has implications for the winds as well as for calculations of gas properties like the optical depth and mass, such as those of \citet{Aspin82, Kemp83, Koch89}. Inclinations derived from gas scattering models are more robust \citep{Berdyugin18}, since they are little effected by gas geometry. However $\Omega$ has some weak dependence on the geometry of the scattering medium \citep{Berdyugin16}, and given a photosphere can be approximated by a shell, an undifferentiated photospheric component could impact other geometrical determinations and hence $\Omega$ to a degree.

\subsection{Inclination}

\begin{figure*}
\includegraphics[width=\textwidth, trim={0.25cm 1.0cm 0cm 0cm},clip]{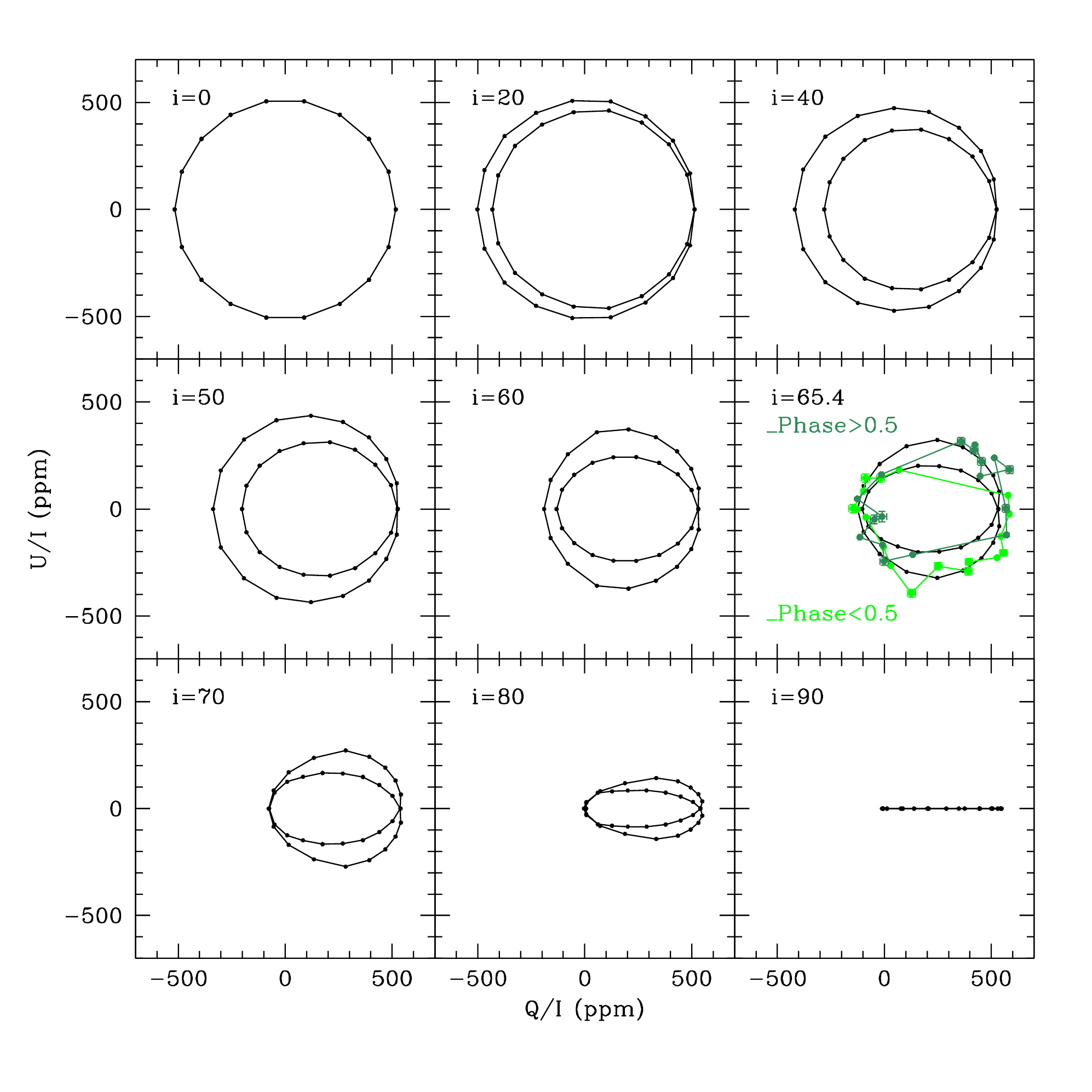}
\caption{Polarization models of $\mu^1$ Sco using the parameters given in Table \ref{tab:modparam} except for the inclination which has been varied from 0 to 90 degrees. Dots are at 10 degree intervals of phase. The $g^{\prime}$ data points are shown on the i = 65.4$\degr$ plot with the dark green points being those for phase > 0.5 that fit the upper of the two intersecting loops, and the light green points being those for phase < 0.5 that fit the lower loop. In this plot the offsets and rotation in Table \ref{tab:fit_pars} have been applied to the observed data points to fit the model, whereas in Fig. \ref{fig:phase} the models were adjusted to fit the data.}
\label{fig:inc}
\end{figure*}

The determination of the inclination of a binary system is important because this information is needed to measure the component masses of a double-lined spectroscopic binary. At present Stellar Population Synthesis models (e.g. \citealp{Eldridge17}) are informed by very few measurements of the heaviest stars. For instance, there are only six O-type stars with mass determinations in DEBCat \citep{Southworth15}. Normally we require for a mass determination a double-lined binary that is also eclipsing, so that the inclination can be measured from the light curve analysis. $\mu^1$ Sco is an example of such a system.

However, as already discussed in section \ref{sec:intro}, polarization observations can provide an alternate means of determining the inclination for non-eclipsing binaries and such methods are described by \citet{Rudy78}, \citet{Brown78} and \citet{Aspin81}. However these studies assume the polarization is due to optically thin circumstellar gas, and the \textit{\mbox{BMcLE} model} in particular, which we no longer think is the correct model for $\mu^1$ Sco, and perhaps other early type binaries.

While $\mu^1$ Sco has a well determined inclination, we explored the effect on our models of varying this parameter. In Fig. \ref{fig:inc} we show the effect on the models (plotted as a \mbox{Q-U} diagram) of changing the inclination value from 0$\degr$ to 90$\degr$ with all other parameters kept at their values in Table \ref{tab:modparam}. This shows a number of interesting features. The largest polarization amplitude is observed for an inclination of 0$\degr$ (i.e. a face-on orbit). For this orientation the degree of polarization is constant, but the polarization position angle rotates through the orbital cycle at a uniform rate. The large polarization in this case, suggests that polarization might be a way of detecting binaries which would not be detected by photometric or spectroscopic methods. Binary frequency increases with mass, and some predictions put the binary fraction for O-type stars at 100 per cent \citep{Sana14}. Consequently, any massive apparently single star is a good target for a polarimetric companion search.

For intermediate inclinations the \mbox{Q-U} diagrams show two intersecting loops corresponding to the two slightly unequal halves of the phase curve. The polarization amplitude varies with inclination and fits the data well for the inclination of 65.4$\degr$ as given in Table \ref{tab:modparam} and used in our earlier models.

For an edge on orbit (i = 90$\degr$) the polarization variation becomes a straight line in the \mbox{Q-U} plane.

We also calculated a set of models with inclination values from 61 to 69$\degr$ and used these to determine the best fitting inclination by fitting the same three parameter model used in Table \ref{tab:fit_pars}. Based on the $g^\prime$ band data we determine a polarimetric inclination of 63.8$\degr$ $\pm$ 2.7, where the uncertainty is based on the same bootstrap analysis used in section \ref{sec:modfit}. This agrees, within the errors, with the inclination of 65.4$\degr$ $\pm$ 1.0 \citep{vanAntwerpen10} and 64.4$\degr$ $\pm$ 0.3 \citep{Budding15} from light curve analysis. The precision of our measurement could probably be improved with additional observations.

\section{Conclusions}
\label{sec:conclusions}

We have discovered phase-locked polarization in the semi-detached eclipsing binary system $\mu^1$~Sco. The initial discovery was made using a small telescope at a city-based observing site with an inexpensive instrument. This demonstrates the accessibility of polarimetry for small institutions and even amateurs.

The phase behaviour seen in $\mu^1$~Sco is complex, and the amplitude of the signal is wavelength dependent, reaching about 700~ppm in green light, and 450~ppm in red light. We fit the polarization phase curves with a SYNSPEC/VLIDORT polarized radiative transfer model that utilises a Wilson-Devinney geometric formalism, and as a result find that most of the polarization arises from photospheric reflection. This adds to an increasing number of other binary systems like u~Her, LZ~Cep, and HD~48099 where it is thought photospheric reflection could play a primary role. The agreement between the model and observations here, along with a similar result for the Spica system, leads us to conclude that polarization by this mechanism must be a common feature of early-type binaries. For more than 50~years scattering by extra-stellar gas has been regarded as the dominant polarigenic mechanism in these systems, but this result begins to suggest a more complicated picture. It will be important to establish to what proportion gas and photospheric mechanisms each contribute in individual systems and in general. To this end we call both for more polarimetric observations of binary systems and further radiative transfer modelling to compare to historical data.

Since polarization by photospheric reflection increases with increasing temperature and reducing surface gravity, the finding has significant implications for the study of massive early-type binary systems. It provides a reliable method for the determination of inclination, and consequently mass, in non-eclipsing double-lined spectroscopic binaries, and encourages a search for missed binary companions in face-on orbits where entrained gas may not be expected. The effect will also need to be accounted for when studying winds in WR+O systems, or other close binaries polarimetrically.

\section*{Acknowledgements}

This research has made use of the SIMBAD database, operated at CDS, Strasbourg, France. This research has made use of NASA's Astrophysics Data System. We thank the former Director of the Australian Astronomical Observatory, Prof. Warrick Couch, and the current Director of Siding Spring Observatory, A/Prof. Chris Lidman for their support of the HIPPI-2 project on the AAT. We thank Prof. Miroslav Filipovic for providing access to the Penrith Observatory, and Darren Maybour for his work on its telescope, systems and assistance when observing. Fiona Lewis, Dag \mbox{Evensberget}, Jinglin Zhao, Shannon Melrose and Rudy Xu also assisted with observations. Funding for the construction of HIPPI-2 was provided by UNSW through the Science Faculty Research Grants Program.

We wish to thank the referee, Andrei Berdyugin, for his contribution to improving the manuscript.

\section*{Data availability}

The primary data underlying this article are available in the article. Auxiliary data is available upon reasonable request.

\bibliographystyle{mnras}
\bibliography{mu1_Sco}


\bsp	
\label{lastpage}
\end{document}